\documentclass{aastex7} % Use for reviewing

\newcommand{\lapprox}{{\lower0.8ex\hbox{$\buildrel <\over\sim$}}}
\newcommand{\gapprox}{{\lower0.8ex\hbox{$\buildrel >\over\sim$}}}
\newcommand{\teff}{\ensuremath{T_{\mbox{\scriptsize eff}}}}
\newcommand{\bpminusrp}{($G_\mathrm{BP} - G_\mathrm{RP}$)}
\newcommand{\gminusk}{($G - K_s$)}
\newcommand{\halpha}{$\mathrm{H}\alpha$}
\newcommand{\llh}{$L_{\mathrm{H}\alpha}/L_{\mathrm{bol}}$}
\newcommand{\llx}{$L_{\mathrm{X}}/L_{\mathrm{bol}}$}
\newcommand{\lbol}{$L_{\mathrm{bol}}$}
\newcommand{\prot}{$P_{\mathrm{rot}}$}
\newcommand{\Ro}{$R_\mathrm{o}$}

\newcommand{\Msun}{\ifmmode {M_{\odot}}\else${M_{\odot}}$\fi}

\newcommand{\columbia}{Department of Astronomy, Columbia University, 550 West 120th Street, New York, NY 10027, USA}
\newcommand{\antioquia}{Instituto de Física, Facultad de Ciencias Exactas y Naturales, Universidad de Antioquia, Medellín, Colombia}

\defcitealias{Newton2017}{N17}

\shorttitle{Anchoring the Chromospheric Activity--Rotation Relation}
\shortauthors{Núñez et al.}
\begin{document}

\title{Anchoring the Chromospheric Activity--Rotation Relation:\\A Study of the Pleiades and Group X} 

\author[orcid=0000-0002-8047-1982]{Alejandro Núñez}
\affiliation{\columbia}
\email[show]{alejo.nunez@gmail.com}  

\author[orcid=0000-0001-7077-3664]{Marcel A. Agüeros} 
\affiliation{\columbia}
\email{m.agueros@columbia.edu}

\author[orcid=0000-0002-2792-134X]{Jason L. Curtis}
\affiliation{\columbia}
\email{jasoncurtis.astro@google.com}

\author[orcid=0009-0004-3949-8894]{Camilo A. Ospinal Achicanoy} 
\affiliation{\antioquia}
\email{camilo.ospinal@udea.edu.co}

\begin{abstract}

We present a study of the rotation and the chromospheric \halpha\ activity of low-mass stars in the $\approx$120 Myr-old Pleiades and the $\approx$400 Myr-old Group X open clusters. We combine published and new rotation periods (\prot) from TESS and ZTF, and use LAMOST, SDSS, and MDM spectra to measure \halpha\ equivalent widths (EWs). While overall the Group~X distributions of \prot\ and EWs are as expected, detailed comparisons to the Pleiades and to the 700-Myr-old benchmark Praesepe produce some surprises. For F through mid K stars, the Group X color--period sequence lies between those of the Pleiades and of Praesepe, as expected. But the \prot\ for its late K and early M dwarfs and the Praesepe slow-rotator sequence overlap, indicating that the 700 Myr-old stars have not spun down significantly despite being nearly twice as old as their Group~X counterparts. In the Rossby number--\llh\ plane, rapid rotators in Group X are saturated, as is seen in older clusters, whereas the rapid rotators in the Pleiades show a weak but measurable dependence of \llh\ on rotation. This implies that the chromospheric activity--rotation relation attains its mature form between 120 and 400 Myr. Dividing the sample by color reveals that K, M0, and M1 dwarfs are no longer saturated by $\approx$120 Myr, M2 and M3 dwarfs become unsaturated between $\approx$400 and $\approx$700 Myr, and later M dwarfs remain largely saturated through field ages even as their saturated \llh\ levels decline, indicating that chromospheric evolution is not determined by rotation period alone.

\end{abstract}


\section{Introduction}

The magnetic fields of low-mass main sequence (MS) stars ($\lapprox$1.3 \Msun) are generated by convective dynamos that couple differential rotation and turbulent motions in their outer envelopes \citep[e.g.,][]{Parker1993}. These magnetic fields heat the stellar atmosphere and drive magnetized winds through which stars lose angular momentum and gradually spin down with age. As rotation slows, magnetic heating and activity decline, producing the link between rotation, activity, and age that underpins both stellar gyrochronology and the age--activity relation \citep[e.g.,][]{Skumanich1972, Noyes1984, Barnes2003, Mamajek2008}. Yet, despite decades of study, the detailed evolution of this activity--rotation--age relation remains incompletely understood.

Activity manifests throughout the stellar atmosphere, from chromospheric \halpha\ and \ion{Ca}{2} H\&K emission to coronal X-rays. These diagnostics consistently show that for large Rossby numbers---defined as the stellar rotation period (\prot) over the convective turnover time $\tau$ \citep[\Ro\ = \prot\ / $\tau$;][]{Noyes1984}---activity declines with increasing \Ro, while for \Ro\ below a threshold \Ro\ it saturates at an approximately constant level \citep[e.g.,][]{Stauffer1994, Randich2000, Pizzolato2003, Wright2011}. The fractional luminosities in \halpha\ and X-rays (\llh\ and \llx, respectively) typically reach saturation near $10^{-4}$--$10^{-3}$, depending on the indicator \citep{Douglas2014, Newton2017, Nunez2024}. At very small \Ro\ (\Ro~$\lapprox$~0.01), several studies have also found a decline in coronal emission, a supersaturation of magnetic activity that may arise from, e.g., centrifugal stripping of the corona \citep{Jardine1999, Jardine2004, Nunez2024}.

Because their members share a common age and composition, open clusters are essential laboratories for exploring the evolution of the activity--rotation relation. When considered together, clusters of different ages trace how stellar dynamos weaken over time. The $\approx$120 Myr-old Pleiades has long served as a benchmark for studying rapidly rotating young stars and the emergence of the saturated regime \citep[e.g.,][]{Rebull2016, Fang2018}. By $\approx$700 Myr, F, G, and K stars in Praesepe and the Hyades are rotating more slowly and have lower overall magnetic activity \citep{Douglas2019, Nunez2024}, while field-age M dwarfs, whose ages are generally unknown but assumed to be several Gyr, show even lower levels of magnetic activity \citep[e.g.,][]{Newton2017}. However, clusters in the 200 to 600~Myr range with comprehensive rotational \textit{and} chromospheric activity measurements are rare, leaving a critical gap in our understanding of how the chromospheric activity--rotation relation evolves between the age of the Pleiades and that of Praesepe.

This age interval is important because it overlaps with the final stages of pre-MS evolution for low-mass stars. At the age of the Pleiades, late K and M dwarfs are still contracting toward the zero-age main sequence \citep[ZAMS;][]{Tognelli2011, Baraffe2015}, so their radii, luminosities, internal structures, and convective properties are still evolving. By the age of Praesepe, much of the early M population has reached the MS, while later M dwarfs may still retain some pre-MS characteristics. Consequently, changes in chromospheric activity between $\approx$120 and $\approx$700 Myr need not reflect rotational spin-down alone: they may also encode the transition from structurally evolving pre-MS stars to MS stars with more stable internal properties.

Finally, the 120--700 Myr interval is central to formulating accurate gyrochronology relations. The rotation period distributions for Praesepe, NGC 6811 ($\approx$1 Gyr old), and NGC 752 ($\approx$1.4 Gyr) have shown that the spin-down of low-mass stars does not follow a smooth, mass-independent braking law, as proposed by \citet{Skumanich1972}. Instead, after converging onto a (relatively) slow-rotating sequence, K dwarfs appear to undergo a temporary epoch of reduced braking efficiency \citep{Agueros2018, Curtis2019}. \cite{Curtis2019} showed that \prot\ for K dwarfs in NGC 6811 overlap those for the same stars in Praesepe despite the $\approx$300 Myr age difference between the two clusters (the former being $\approx$40\% older than the latter), and argued that the duration of this phase of stalled spin-down increases as mass decreases. 

\citet{Curtis2020} used \prot\ measurements in the $\approx$2.5 Gyr-old NGC 6819  and Ruprecht 147 open clusters to show that these stalled stars eventually resume spinning down by $\approx$2.5 Gyr and that the resulting color--period sequence becomes relatively flat. When this phase of stalled spin-down for K and M dwarfs begins, however, remains less constrained.

The open cluster Group~X occupies this intermediate-age regime, with an estimated age of $\approx$300--400 Myr \citep{Tang2019, Fürnkranz2019, Messina2022}. It therefore probes the epoch when rotational evolution, magnetic braking, and low-mass stellar contraction are all changing simultaneously. Several recent studies have established high-quality Gaia-based membership catalogs for Group~X \citep[e.g.,][]{Tang2019, Olivares2023, Hunt2024} and mapped its rotation distribution in detail \citep{Messina2022, Newton2022}. Yet, the magnetic behavior of its members is largely unexplored.

We present a comprehensive characterization of the chromospheric activity--rotation relation for members of Group~X, combining new and archival optical spectra and new and published rotation periods to analyze how \halpha\ emission depends on rotation in this cluster. We also consolidate published \halpha\ and rotational data for the Pleiades. By comparing our results for both clusters to those of the older Praesepe and field M dwarfs, we aim to trace the evolution of the chromospheric activity--rotation relation over around an order of magnitude in age.

The paper is organized as follows: Section \ref{sec:members} describes the brief history of Group~X and the membership catalogs we used for Group X and the Pleiades, along with our efforts to identify binaries in both clusters. Section \ref{sec:prots} presents the rotational data for both Group~X and the Pleiades. Section \ref{sec:halpha} summarizes the spectroscopic data and \halpha\ measurements for both clusters. Section \ref{sec:results} presents the rotational, chromospheric, and activity--rotation analyses and comparisons between the Pleiades, Group~X, Praesepe, and field M dwarfs. We conclude in Section \ref{sec:conclusion}.

\section{The Clusters, Their Properties, and Those of Their Members} \label{sec:members}

\subsection{What's in a Name? The Evolving Identity of Group~X}
The stellar aggregate now known as Group~X was first identified by \citet{Oh2017}, who applied an unsupervised clustering algorithm to the Tycho--Gaia Astrometric Solution data, part of Gaia Data Release 1 \citep[][]{GaiaDR1, Lindegren2016}, and assigned sequential numbers to the resulting comoving associations. In that catalog, the structure corresponding to Group~X appeared as ``group 10''---the numerical label serving only as an index rather than a name.

\citet{Faherty2018} confirmed that group 10 was a previously unknown, nearby open cluster and, using Gaia Data Release 2 \citep[DR2;][]{GaiaDR2}, expanded its membership list while retaining the same provisional designation. Soon after, two independent studies, by \citet{Tang2019} and by \citet{Fürnkranz2019}, rediscovered the  structure in Gaia DR2. \citet{Tang2019} recognized its equivalence to group 10 from \citet{Oh2017} but, without explanation, renamed it ``group-X.'' \citet{Fürnkranz2019} did not assign a name to the (re)discovered cluster.

\citet{Sapozhnikov2021} subsequently confirmed that group-X and the nearby Coma Berenices cluster are distinct, and retained the nomenclature from \citet{Tang2019} in the slightly different form Group~X. \citet{Newton2022} used the hyphenated form Group-X in a study of rotation in the cluster, whereas \citet{Messina2022}, who also surveyed the cluster for rotation, adopted the unhyphenated Group~X, a convention that has gained wider acceptance in recent literature.

Further work using Gaia Early Data Release~3 \citep[EDR3;][]{edr3} revealed possible substructure in Group~X. \citet{He2022} identified two components---named Group~X-a and Group~X-b---with nearly identical ages, but slightly offset positions and velocities. \citet{Olivares2023} independently reached a similar conclusion, proposing that Group~X is really Latyshev~2 and Mecayotl~1, the first known and the second new. Latyshev 2 is an open cluster introduced by \citet{Latyshev1977} and later cataloged by \citet{Archinal2003}. That seven-star association was later deemed non-physical \citep{Mamajek2016}, but three of its members belong to Group~X, and \citet{Olivares2023} therefore relabeled the modern Gaia structure Latyshev 2. These authors then introduced Mecayotl~1\footnote{\textit{Mecayotl} is the Nauhatl word for network and is the name of the semi-supervised clustering algorithm developed by \citet{Olivares2023} to identify stellar associations.} as a nearby loose stellar association with $\approx$150 members. Although the derived kinematic differences between the two structures are modest, this dual-component picture broadly echoes the findings by \citet{He2022}.

Finally, \citet{Hunt2023} applied the HDBSCAN clustering algorithm to Gaia Data Release 3 \cite[DR3;][]{GaiaDR3} data and compiled a catalog of open clusters. Group~X appears as HSC 759 (``HDBSCAN Star Cluster 759''); the authors did not cross-match HSC~759 with previously named associations.

The unhyphenated, capitalized ``Group~X'' most faithfully reflects the introduction of the name for this cluster by \citet{Tang2019} and its subsequent use. To avoid further fragmentation of terminology, we recommend that future studies adopt Group~X as the standard designation for this cluster. It is, after all, a name with an appropriately modern backstory: unlike well-known open clusters whose designations hark back to Antiquity (e.g., Hyades, Pleiades), to individual catalogers (e.g., Melotte 20, Messier 37), or to the early photographic surveys that numbered them (e.g., IC 2391, NGC 752), Group~X emerged from the era of automated mining of Gaia catalogs. The result is a designation that reflects both the data-driven nature of modern astronomy and the sense of discovery that continues to shape our understanding of nearby stellar populations.

\subsection{Assembling Cluster Membership Catalogs} \label{sec:bonafide}
To create our catalog of Group~X members, we consolidated several Gaia-based membership catalogs. We started with those of \citet{Fürnkranz2019} and \citet{Tang2019}, who found 177 and 218 Group X members, respectively, using Gaia DR2 data. Merging the two catalogs produced a list of 239 stars that are present in one or both of these studies.

We then added the catalog of \citet{Newton2022}, who used Gaia EDR3 to identify 198 additional members, of which 105 were considered high probability. We also added the catalog of \citet{Olivares2023}, who identified 88 members with EDR3 that were not included in any previous studies. Finally, we added the catalog of \citet{Hunt2024}, who added 41 members using Gaia DR3.  

\begin{figure}
\centerline{\includegraphics{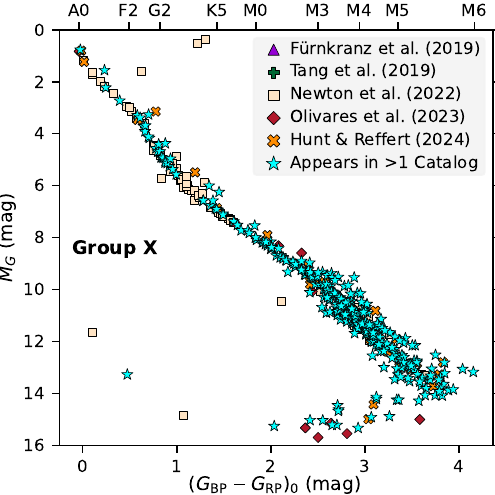}}
\vspace{-0.2cm}
\caption{Gaia CMD for the 487 stars in our Group~X catalog, a consolidation of several Gaia-based catalogs in the literature. Spectral types are indicated along the top axis for reference. The different symbols and colors indicate the provenance of each star. Stars present in more than one catalog (72\% of the total) are indicated with cyan star symbols.}
\label{fig:cmdprovenance}
\end{figure}

This merging produced a list of 487 stars, 350 of which appear in more than one of the catalogs we considered. In Figure \ref{fig:cmdprovenance}, we  show the resulting color--magnitude diagram (CMD) where we color-code the points by the catalog in which they were originally listed, calculate absolute $G$ magnitudes ($M_G$) using Gaia parallaxes, and provide approximate spectral types for reference.\footnote{\label{note} We used the E.~Mamajek table (version 2022.04.16; \url{http://www.pas.rochester.edu/~emamajek/EEM_dwarf_UBVIJHK_colors_Teff.txt}) to estimate spectral types from Gaia colors. Much of the table is from \citet{Pecaut2013}.}

To create our catalog of Pleiades members, we used the following Gaia-based catalogs: \citet[][based on DR2]{Cantat-Gaudin2018}, \citet[][EDR3]{Torres2021}, \citet[][EDR3]{Heyl2022}, \citet[][DR3]{Alfonso2023}, and \citet[][DR3]{Hunt2024}. %We did not consider pre-Gaia catalogs as they have been superseded by Gaia-based work. 
Combining these catalogs produced a total of 994 cluster stars.  

All stars in the Gaia-based Pleiades membership catalogs of \citet{GaiaCol2018}, \citet{Lodieu2019}, and \citet{vanGroeningen2023} are already included in the five catalogs we used. In addition, only eight of the 994 stars in our combined catalog are \textit{not} in the \citet{Hunt2024} catalog. Our final Pleiades catalog, therefore, is almost completely defined by stars from this latter study. Table \ref{tbl:catalog} describes the columns in our full catalog, available online.

\subsection{Assigning Cluster Ages}\label{sec:ages}

We adopted an age of 400 Myr for Group~X. This age was derived by \citet{Fürnkranz2019} and \citet{Tang2019} from isochrones. This age also agrees with that derived by \citet{Sapozhnikov2021}, 450$\pm$100 Myr, also from isochrones, and is in rough agreement with that derived by \citet{Messina2022}, 300$\pm$60 Myr, who compared the cluster's \prot\ distribution to that of older and younger clusters. For all Group~X members, we assumed zero reddening.
% and [Fe/H] = 0.0, as noted by \citet{Tang2019}
%Tang2019: 300-400 Myr
%Fürnkranz2019: 400 Myr
%Newton2022: 300+-50 Myr (adopted from NGC 3532 Fritzewski2019)
%Messina2022: 300+-60 Myr (gyrochronology)
%Olivares2023: 300 Myr (Latyshev 2) and 400-600 Myr (Mecayotl 1)
%Hunt2024: 149+127-73 Myr (derived from OC density in MW; unreliable)

We adopted an age of 120 Myr for the Pleiades. This value agrees with age estimates derived by \citet{Stauffer1998} and \citet{Galindo-Guil2022} (by analyzing the cluster's lithium depletion boundary) and by \citet{Frasca2025} (by fitting empirical lithium depletion isochrones).

For all Pleiads, we assumed E($B-V$) = 0.04 \citep{GaiaCol2018}. Although reddening is known to vary across the Pleiades, with some members more heavily reddened than the cluster mean \citep[e.g.,][]{Soderblom1993}, the stellar quantities used in our analysis are derived primarily from Two Micron All Sky Survey \citep[2MASS;][]{2mass} $K_s$ photometry, which is much less affected by interstellar extinction than optical bands. We nevertheless tested the effect of adopting larger individual reddening values and found that the resulting changes in the quantities used in our analysis are at most $\approx$9\%, with negligible impact on our results.

\begin{deluxetable}{@{}ll@{}}
\digitalasset
\tabletypesize{\footnotesize} 

\tablecaption{Columns in our Clusters membership catalog \label{tbl:catalog}}

\tablehead{
\colhead{Column} & \colhead{Description}
}

\startdata
1      & Gaia DR3 source identifier \\
2      & 2MASS designation \\
3      & Open cluster \\
4, 5   & R.A., Decl. for epoch J2000\\
6, 7   & Gaia DR3 $G$ magnitude and 1$\sigma$ uncertainty \\
8      & Gaia DR3 \bpminusrp \\
9, 10  & 2MASS $K_s$  magnitude and 1$\sigma$ uncertainty \\
11, 12 & Distance and 1$\sigma$ uncertainty \\
13     & Binary? \\
14     & Candidate photometric binary? \\
15     & Gaia DR3 RUWE \\
16, 17 & Effective temperature \teff\ and 1$\sigma$ uncertainty \\
18, 19 & \lbol\ and 1$\sigma$ uncertainty \\
20, 21 & Stellar mass $M_*$ and 1$\sigma$ uncertainty \\
22     & Rotation period \prot \\
23     & Source of \prot\tablenotemark{a} \\
24     & Convective turnover time $\tau$ \\
25     & Rossby number \Ro \\
26, 27 & Measured \halpha\ equivalent width EW and 1$\sigma$ uncertainty\\
28     & Relative \halpha\ EW \\
29, 30 & $\chi$ and 1$\sigma$ uncertainty \\
31, 32 & \llh\ and 1$\sigma$ uncertainty \\
33     & Spectral type derived from ($G - K$)$_0$ \\
\enddata
\tablecomments{The contents of this table is published in its entirety in the electronic edition of the Journal.}
\tablenotetext{a}{C16: \citet{Covey2016}; F25: \citet{Frasca2025}; H10: \citet{Hartman2010}; M22: \citet{Messina2022}; N22: \citet{Newton2022}; R16: \citet{Rebull2016}; TARS: \citet{TARS}; TZ: new TESS and/or ZTF.} 

\vspace{-0.25in}

\end{deluxetable}

%4      & Membership classification\tablenotemark{a} \\
%7, 8   & Gaia DR3 $\mu_\alpha$ cos $\delta$ and 1$\sigma$ uncertainty \\
%9, 10  & Gaia DR3 $\mu_\delta$ and 1$\sigma$ uncertainty \\
%13, 14 & Gaia DR3 Radial velocity and 1$\sigma$ uncertainty \\
%15--17 & Cartesian U, V, and W velocities\tablenotemark{a}\\

\subsection{Flagging Binaries} \label{sec:bin}

To identify candidate binaries, we flagged stars with Gaia Renormalized Unit Weight Error (RUWE) $>$1.4, as a large RUWE is commonly associated with unresolved companions \citep[e.g.,][]{Belokurov2020, Stassun2021}. We also cross-matched our sample with Gaia's Non-Single Star (NSS) catalog \citep{GaiaDR3}. And we flagged stars with Gaia Image Parameter Determination (IPD) multipeak fraction $>$ 4\% (\texttt{ipd\_frac\_multi\_peak} $>$ 4). In their study of unresolved binarity in the Pleiades, \citet{Chulkov2025} found that the multipeak fraction parameter above this value was a strong predictor of subarcsecond multiplicity.

\begin{figure*}
\centerline{\includegraphics{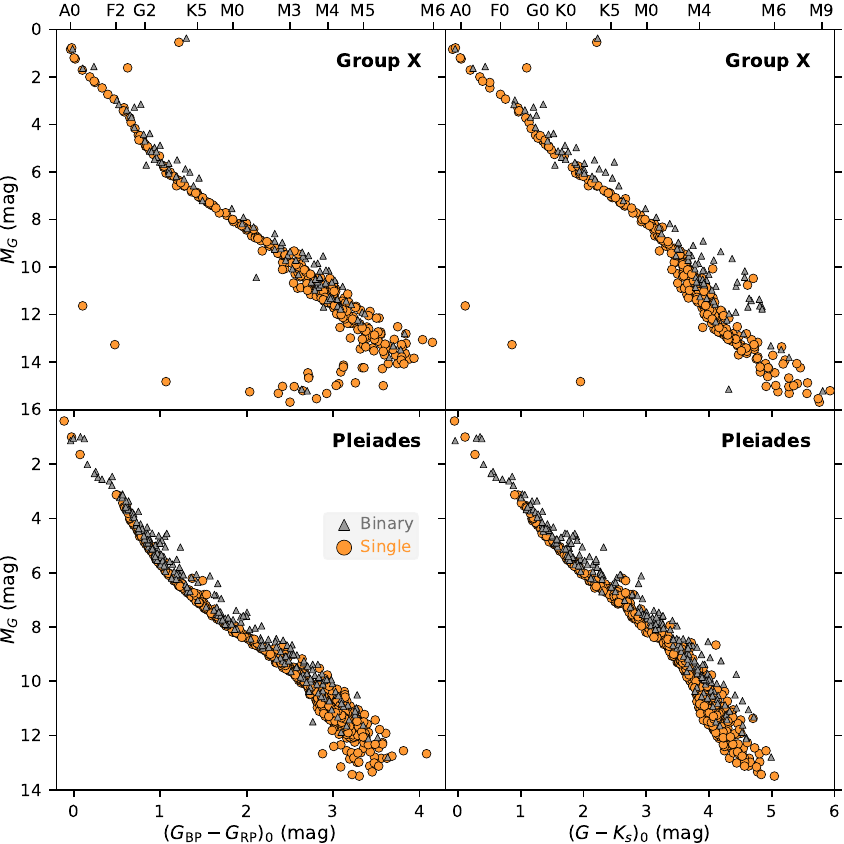}}
\vspace{-0.1cm}
\caption{\bpminusrp$_0$ (left column) and \gminusk$_0$ (right column) CMDs for Group~X (top row) and Pleiades (bottom row) members. The filled gray triangles are stars flagged as candidate or known binaries, as described in Section~\ref{sec:bin}. }
\label{fig:cmdbins}
\end{figure*}

In addition, we flagged binaries reported in the literature, beginning with four known Group X eclipsing binaries: \object{Gaia DR3 1395494252395031552} \citep{Drake2014}, \object[Gaia DR3 1503770755884281344]{1503770755884281344} \citep{Hartman2011}, and \object[Gaia DR3 1584856779198650112]{1584856779198650112} and \object[Gaia DR3 1608637983053293696]{1608637983053293696} \citep{Chen2020}. There is also a known triple system, cataloged in \citet{Tokovinin2018} as 13232+3351, consisting of \object{Gaia DR3 1469850341130739840}, \object[Gaia DR3 1469850341130739712]{1469850341130739712}, and \object[Gaia DR3 1469850341130727040]{1469850341130727040}. For the Pleiades, we used binary classifications from \citet{Stauffer1987}, \citet{Soderblom1993}, \citet{Queloz1998}, \citet{Fang2018}, and especially \citet{Torres2021}, which describes an extensive long-term spectroscopic survey of Pleiads.

To identify candidate unresolved photometric binaries, we used a Gaia $G$ minus 2MASS $K_s$ CMD, adopting a 400~Myr and a 120 Myr PAdova and TRieste Stellar Evolution Code (PARSEC) isochrone \citep{bressan2012} as reference for the single-star MS in Group X and the Pleiades, respectively. A star was flagged as a photometric binary candidate if it sits more than 0.375 mag above the isochrone, which corresponds to the midpoint between the single-star sequence and the expected locus of equal-mass binaries \citep[0.75 mag above the MS; see][]{Steele1995}. We applied this binary identification to stars with $0 <$ \gminusk$_0$ $< 3.2$ ($\approx$A0 to $\approx$M1) for Group X and $0 <$ \gminusk$_0$ $< 2.5$ ($\approx$A0 to $\approx$K5) for the Pleiades, restricting the analysis to the color ranges over which stars are expected to have reached the MS at the respective cluster ages \citep[e.g.,][]{Baraffe2015, Somers2017}.

In Group X we found 21 candidate photometric binaries, including 12 with RUWE $>$ 1.4, the star in the NSS catalog with RUWE $<$ 1.4, and seven stars with IPD fraction $>$ 4\%. In the Pleiades, we found 65, including 27 with RUWE $>$ 1.4, one star in the NSS catalog with RUWE $<$ 1.4, and six stars with IPD fraction $>$ 4\%.

In Table \ref{tbl:catalog}, we identify stars that have at least one of these various binary flags, and we also identify stars that we flagged as candidate photometric binaries. In all, $\approx$22\% of Group~X and $\approx$22\% of Pleiades stars are either candidate or known binaries.

We combine our membership and binarity analyses in Figure \ref{fig:cmdbins}, which shows the CMD of Group~X (top row) and of the Pleiades (bottom row) in both Gaia \bpminusrp$_0$ color (left column) and in \gminusk$_0$ color (right column) and highlights the stars flagged as binaries.

\subsection{Deriving Stellar Properties}\label{sec:properties}
We derived fundamental stellar parameters for members of both clusters from near-infrared photometry\footnote{Near-infrared photometry provides a metallicity-insensitive proxy for fundamental parameters for low-mass stars \citep[e.g.,][]{Baraffe1998, Delfosse2000, Mann2015}.} combined with Gaia-based distances. For the latter, we adopted individual probabilistic distances from \citet{Bailer-Jones2021}, which are based on Gaia EDR3 parallaxes. The median distance of Group~X members is 101 pc, and of Pleiads, 136 pc.

We cross-matched the Gaia DR3 catalog with 2MASS to obtain $K_s$ magnitudes for our sample of cluster stars. In Group~X, five stars did not have a 2MASS counterpart, and in 19 cases Gaia resolved two sources that 2MASS treated as a single object, resulting in a duplicate association. In addition, seven stars had unreliable 2MASS photometry (quality flag E, F, or U). In the Pleiades, there were 11 cases of duplicate association and five stars with unreliable 2MASS photometry. For all these cases (no 2MASS counterpart, duplicate association, or unreliable 2MASS photometry), we estimated $K_s$ magnitudes using the Gaia--2MASS photometric transformations of \citet{Riello2021}.

Next, we computed absolute $K_s$ magnitudes ($M_K$), which were then translated into effective temperatures (\teff), bolometric luminosities (\lbol), and stellar masses ($M_\star$) by interpolating within the empirical relations tabulated by E.~Mamajek.\footnote{See Footnote~\ref{note}.} These stellar quantities are included in our final catalog (see Table \ref{tbl:catalog}).

Propagating the $K_s$ photometric and distance uncertainties results in \teff\ uncertainties of $<$5\% throughout our sample. As an external check, we compared these photometric temperatures with spectroscopic temperatures from the Large Sky Area Multi-Object Fiber Spectroscopic Telescope \citep[LAMOST;][]{lamost} Data Release 11 Low Resolution Catalog.\footnote{\url{https://www.lamost.org/dr11/v2.0/}} We found LAMOST temperatures to be typically $\approx$100 to 300 K cooler for stars with \teff\ $\lapprox$ 3800 K and $\approx$200 to 300 K hotter for stars with \teff\ $\gapprox$ 4300 K. We also tested the potential effect of starspot-induced dimming by increasing the adopted $K_s$ magnitudes by 0.1 mag \citep[see, e.g.,][]{Gully-Santiago2017, Cao2022}; the resulting \teff\ values changed by a median of only $\approx$1\% and by at most $\approx$4\%. These tests indicate that the $M_K$-based stellar parameters are sufficiently robust for the purposes of our analysis.

As with other photometry-based methods, unresolved multiplicity will bias the results: binarity causes stars to appear brighter than single counterparts, leading to systematically overestimated \teff, \lbol, and $M_\star$ values. We therefore did not include confirmed or candidate binaries when performing analyses that involved these quantities.

\section{Assembling Rotation Period Catalogs} \label{sec:prots}

\subsection{From the Literature} \label{sec:prots_lit}

\subsubsection{Group X}

\citet{Messina2022} and \citet{Newton2022} used Transiting Exoplanet Survey Satellite \citep[TESS;][]{Ricker2015} light curves to measure \prot\ values for members of Group~X. The latter authors also used Zwicky Transient Facility \citep[ZTF;][]{masci2018} data to complement and/or confirm TESS-derived periods for stars observed by both TESS and ZTF.

\begin{figure}
\centerline{\includegraphics{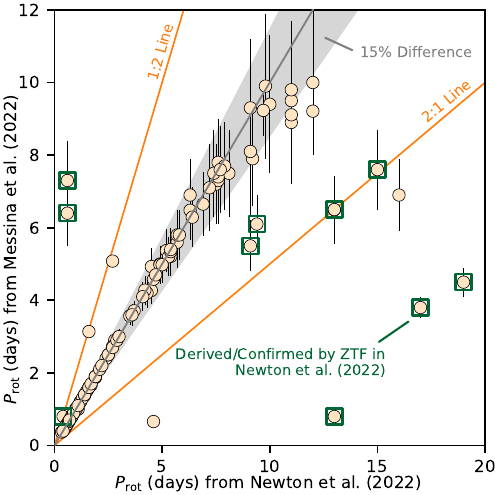}}
\vspace{-0.2cm}
\caption{\prot\ from \citet{Newton2022} vs.~\prot\ from \citet{Messina2022} for the 118 Group~X members with a \prot\ measurement in both studies. The solid gray line is the 1:1 match, and the shaded area is the $\pm$15\% difference range. The orange lines are the 1:2 and 2:1 lines. For stars outside the $\pm$15\% range, we highlight with green squares those that were derived from or confirmed by ZTF light curves in \citet{Newton2022}. The overall agreement between the measurements is excellent, with a median difference $<$2\%.}
\label{fig:protcomp}
\end{figure}

%We consolidated \prot\ values from both studies.
There are 118 Group~X members with \prot\ measurements in both studies. %(only \citealt{Messina2022} estimated \prot\ uncertainties). 
In Figure \ref{fig:protcomp}, we compare \prot\ values from both studies. Most measurements fall close to the one-to-one line, and for the vast majority of stars the two measurements agree to within 15\%.\footnote{Only \citealt{Messina2022} estimated \prot\ uncertainties, with a median uncertainty value of 4\%.} For those that disagree by more than 15\%, we indicate with green squares \prot\ values from \citet{Newton2022} that were derived solely from ZTF or for which TESS- and ZTF-derived period measurements agreed. Due to the differences in sampling and duration of the TESS and ZTF light curves, \prot\ values that agree in both are much more reliable than those derived only from TESS. For our catalog, we adopted the \prot\ values from \citet{Newton2022} for stars that also have a \prot\ in \citet{Messina2022}. 

More recently, \cite{TARS} published the TESS All-Sky Rotation Survey (TARS) catalog, in which periods were measured for $>$10$^6$ stars within 500~pc. These authors used a methodology that can correct for half-period aliases in the \prot\ and that can identify systematics that inject spurious signals from scattered light into the light curves---such phenomena can mimic rotational signals in periodogram analyses. In these authors' version 2 catalog, 249 Group X members have periods, of which 72 are stars that previously did not have a \prot. 

Only 10 TARS \prot\ disagree from those in \citet{Newton2022} and/or \citet{Messina2022} by $>$$5\%$. In these 10 cases, we adopted the TARS values based on the more extensive TESS sector coverage of the TARS catalog. Combining all these measurements, we have a total of 286 Group X stars with a \prot.

\subsubsection{The Pleiades}\label{sec:prots_pl}

We consolidated \prot\ measurements obtained from the Hungarian-made Automated Telescope Network (HATNet) data \citep{Hartman2010}, K2 data \citep{Rebull2016}, Palomar Transient Factory (PTF) data \citep{Covey2016}, TESS data \citep{Frasca2025}, and from the TARS catalog. We ignored seven \prot\ included in \citet{Hartman2010} that came from the literature and  were considered uncertain in the original studies. From the TARS catalog, we obtained 382 \prot\ values, of which 19 are for stars that did not have previous \prot\ measurements. In total, we have 950 Pleiads---929 of which are low-mass stars---with at least one \prot\ measurement.

\begin{figure*}
\centerline{\includegraphics{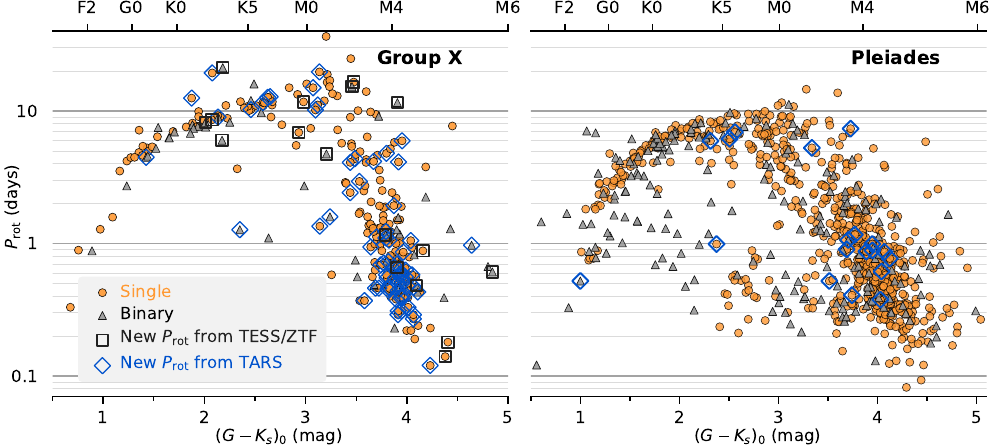}}
\vspace{-0.2cm}
\caption{Color vs.~\prot\ for single members (orange circles) and binary members (gray triangles) of Group X (left panel) and of the Pleiades (right panel). Our new \prot\ measurements from TESS/ZTF data for Group X stars are highlighted with black squares. New \prot\ measurements from the TARS catalog are highlighted with blue diamonds.}
\label{fig:prot}
\end{figure*}

There are 540 Pleiads with two or more \prot\ values. In the vast majority of cases, the \prot\ values were within 15\% of one another, and we adopted the mean of the values as the \prot\ in our catalog.

There are 39 stars with \prot\ values discrepant by $>$15\%. Seventeen of these cases required careful consideration to decide which \prot\ to trust. Appendix \ref{app:prots} describes in detail how we approached these 39 cases and our final \prot\ selection for them.

\subsection{Adding New Rotation Periods in Group X} \label{sec:prots_new}

To expand the rotational census of Group X, we focused our efforts on cluster members with \halpha\ measurement, using methods identical to those applied in \citet{Agueros2025}. Briefly: we constructed TESS and ZTF light curves for these stars. We downloaded 40$\times$40 pixel cutouts from TESS Full-Frame Images \citep{TESSdata} using \texttt{TESScut} \citep{TESScut}, and generated light curves with Causal Pixel Modeling \citep[][]{wang2016}, as implemented in \texttt{unpopular} \citep{Hattori2022}. We derived periods with Lomb-Scargle periodograms, and validated \prot\ visually sector by sector. Because individual TESS sectors span $\approx$27~days and include mid-sector data gaps, our search was limited to \prot\ $\lapprox$ 15 days. When multiple sectors were available, we used them to confirm the stability and robustness of the measured period. In this manner, we found periods for 49 stars.

We complemented these space-based data using multi-year ZTF light curves, which provide a much longer time baseline but at lower cadence and higher noise. Following the procedure in \citet{Agueros2025}, we extracted differential photometry relative to nearby reference stars and computed Lomb--Scargle periodograms. Although sparser than TESS, the ZTF light curves are particularly valuable for confirming TESS-derived periods and enabling measurements beyond the TESS de facto $\approx$15-day limit. In this manner, we found periods for 31 stars.

Several stars for which we measured a TESS and/or a ZTF \prot\ also appear in the TARS catalog. In six cases we adopted our own TESS/ZTF \prot\ value over the value in the TARS catalog, after identifying potential harmonics or stronger Lomb-Scargle power signals. %We identified new TESS/ZTF \prot\ values for 21 stars in Group X. 
For cluster members with both TESS- and ZTF-derived periods, we adopted the average of the two as our catalog \prot\ value when the periods agreed within 10\%. For \prot\ $>$ 10 days, we instead adopted the ZTF value, as those light curves are better suited for measuring longer periods.

Combining our new measurements from those in the literature, we have 306 stars in Group X with a \prot. Of these, we identified 18 \prot\ measurements as likely being impacted by blending from neighboring stars. For most of the 18 stars, the neighbors are similar in color and magnitude and too close to permit localizing the signal to one or the other star. For three cases in which one neighbor is significantly fainter than the other, we assigned the shared period to the brighter star and maintained that \prot\ value in our catalog. Our final Group~X \prot\ sample is therefore 291 stars.

With the new TARS \prot\ values and our new TESS/ZTF-derived \prot\ measurements, we not only increased by $\approx$30\% the number of Group~X stars with rotational data, we also refined the existing data by better identifying blended or harmonic cases. The result is an improved sampling of the Group~X color--period sequence, especially among M2 to M4 type dwarfs. We include all \prot\ values for Group X and the Pleiades and their origin in our final catalog (see Table~\ref{tbl:catalog}). Figure~\ref{fig:prot} displays new \prot\ measurements from our own analysis (black squares) or from TARS (blue diamonds) along with values from the literature for both clusters. Combined with our updated binary flags for both clusters, our data provide an updated view of the 120-Myr and 400-Myr color--period distributions.

\subsection{Calculating Rossby Numbers}\label{sec:ro}
The Rossby number \Ro\ provides an empirically motivated way of normalizing rotation, such that stars of different masses but similar \Ro\ tend to exhibit comparable levels of magnetic activity \citep{Noyes1984}.
%We estimated $\tau$ values as a function of $M_\star$ using the prescription of \citet{Wright2018}.
We followed \citet{Wright2018} in using mass to estimate the convective turnover time $\tau$. For each star in our sample, we interpolated log($\tau$) from these authors' tabulated relation, adopting the appropriate value given the star’s $M_\star$. Our $\tau$ and \Ro\ values for Group~X and Pleiades stars are included in the online catalog (see Table \ref{tbl:catalog}).

\section{Chromospheric Activity With \halpha}\label{sec:halpha}

\subsection{Optical Spectroscopy} \label{sec:spectra}
\subsubsection{From Our MDM Observations} \label{sec:mdm}
We obtained spectra of Group~X stars over 12 observing runs between 2021 September and 2026 March with the Ohio State Multi-Object Spectrograph (OSMOS) on the MDM Observatory 2.4-m Hiltner telescope (see \autoref{tbl:mdm}). We used the blue 4K detector (OSMOS 4K) with a 1$\farcs$2 inner slit, for an approximate wavelength coverage of 4000--6800 \AA, $\approx$0.7 \AA\ pixel$^{-1}$ sampling, a resolution $R$~$\approx$~1600, and peak efficiency at 6400~\AA.

\begin{deluxetable}{@{}lc@{}}
\centering 
\tabletypesize{\normalsize}

\tablecaption{MDM observations of Group X
 \label{tbl:mdm}}

\tablehead{
\colhead{Date} &
\colhead{\# of Spectra}
}

\startdata
2021 Sep 04          &  2 \\
2022 Mar 26$-$Mar 31 & 36 \\
2022 Apr 21$-$Apr 23 & 38 \\
2022 May 13          & 13 \\
2023 Mar 24$-$Apr 01 & 30 \\
2024 Mar 28$-$Mar 31 & 27 \\
2024 Apr 17$-$Apr 22 & 13 \\
2024 May 08$-$May 09 &  4 \\
2024 Jun 05$-$Jun 07 & 10 \\
2025 May 03$-$May 07 & 28 \\
2026 Feb 05$-$Feb 10 & 12 \\
2026 Mar 12$-$Mar 28 &  9 \\
\hline
Total & 222
\enddata
\tablecomments{All dates are UT.}

\end{deluxetable}

We collected 222 spectra for 207 Group~X members. We reduced our MDM spectra with the Python package {\tt PypeIt} \citep{pypeit, pypeit_zenodo}. All the spectra were trimmed, overscan- and bias-corrected, cleaned of cosmic rays, flat-fielded, extracted, dispersion-corrected, and flux-calibrated. The median signal-to-noise ($S$/$N$) for our  spectra is 81 at \halpha. The reduced MDM spectra are archived online.\footnote{Available from the Columbia University Academic Commons under a CC0 license: \url{https://doi.org/10.7916/g2wg-y563}.}

\subsubsection{From the SDSS and LAMOST Archives} \label{sec:lamost_sdss}
We searched the Sloan Digital Sky Survey \cite[SDSS;][]{york00} science archive web server\footnote{\url{https://dr19.sdss.org/zora/}; last consulted 2025 July 26.} for optical spectra in the SDSS Data Release 19 \citep{sdss_dr19}. SDSS spectra are sky-subtracted, corrected for telluric absorption, spectrophotometrically calibrated, and provided in heliocentric vacuum wavelengths. The wavelength coverage is 3600--10400~\AA\ with $R$~$\approx$~2000 at \halpha\ \citep{sdss_spec}. After excluding spectra with $S/N \lesssim 5$, we retained SDSS spectra for 48 Group~X stars.

We also searched the LAMOST Data Release 11 Low Resolution Catalog. LAMOST low resolution spectra are wavelength- and flux-calibrated and sky-subtracted, with coverage of 3690--9100~\AA\ and $R$~$\approx$~1800 at 5500~\AA. After applying the same $S/N \lesssim 5$ cut, we retained spectra for 118 Group~X stars and for 612 Pleiads.

In total, we assembled 429 optical spectra of 323 Group X members. %, drawing on our MDM campaign as well the archival SDSS and LAMOST data. 
We have at least one spectrum for 65\% of the low-mass stars in Group~X and for 63\% of low-mass Pleiads.

\subsubsection{Measuring \halpha\ Equivalent Widths} \label{sec:ew}

\begin{figure*}
\centerline{\includegraphics{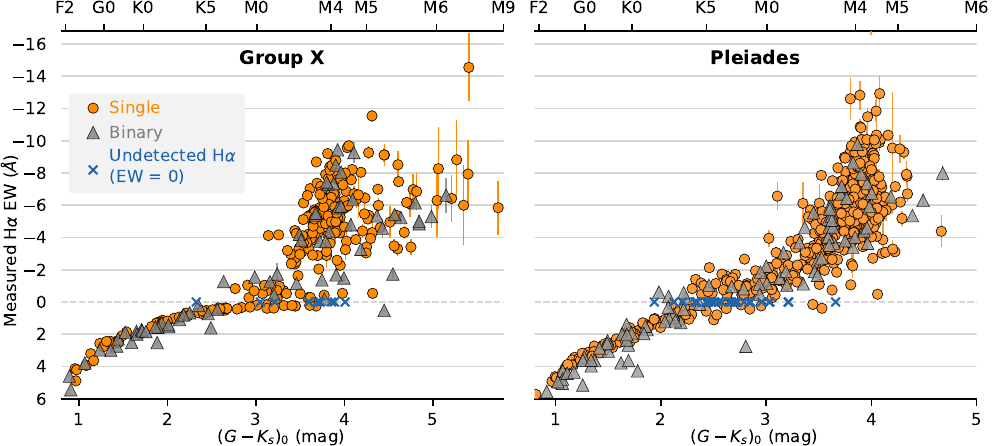}}
\vspace{-0.2cm}
\caption{Color vs.~measured \halpha\ EW for Group~X members (left panel) and Pleiads (right). Negative EW values represent emission. Single stars are indicated with orange circles and binaries with gray triangles. Most of the EW error bars are smaller than the symbols. For figure clarity, we excluded one outlier Group X star with \gminusk$_0$ = 5.3 mag and EW = $-22$ \AA\ and one Pleiad with \gminusk$_0$ = 4.0 mag and EW = $-17$ \AA, although we included these two measurements in our subsequent analyses. We excluded from our study stars with \gminusk$_0$ $<$ 0.8 mag (spectral types earlier than $\approx$F2), as they are not relevant to our analysis. Blue $\times$ symbols indicate stars for which \halpha\ is undetected in the spectra, and for which we set EW = 0 \AA. The EWs shown here were not corrected for the quiescent \halpha\ absorption present in these stars.}
\label{fig:ew}
\end{figure*}

We measured the equivalent width (EW) of the \halpha\ line in each spectrum using \texttt{PHEW} \citep{phew}, which provides an automated interface to \texttt{PySpecKit} \citep{pyspeckit}. We selected continuum windows bracketing \halpha\ that typically spanned tens of \AA\ and were customized for each spectrum and fitted and integrated the line profile after local normalization. \texttt{PHEW} assesses uncertainties through a Monte Carlo approach: 1000 realizations of each spectrum are generated by perturbing the fluxes according to their errors. We adopted the dispersion of the resulting EW distribution as the formal $1\sigma$ error.

For stars observed more than once, we combined the individual measurements into an error-weighted mean EW and associated uncertainty. By convention, negative values indicate emission, and a value of exactly zero indicates that no \halpha\ feature could be detected at our spectral resolution. We excluded stars in the latter category from our subsequent analysis, as we did not measure an uncertainty around their assigned EW = 0 \AA\ values.

Figure \ref{fig:ew} shows the \halpha\ EW distribution for low-mass stars in Group~X (left panel) and the Pleiades (right), with orange circles representing single members, and gray triangles, binaries. We highlight with blue crosses the stars for which we could not detect the \halpha\ line.

\citet{Fang2018} also used LAMOST low-resolution spectra to measure \halpha\ EWs for a sample of Pleiads. There are 238 stars in that study for which we have an EW measurement. For the 162 stars with \halpha\ in emission, our EW measurements are usually within $\approx$20\% of their values. On the other hand, for the 76 stars with \halpha\ in absorption, our EW measurements are between 30\% and 60\% larger than those in \citet{Fang2018}. This systematic discrepancy may be the result of the automated approach in \citet{Fang2018} in defining the wavelength range of the continuum around the \halpha\ line: the typical \halpha\ absorption line profile has broad tails that can span up to $\approx$20~\AA\ on both sides of the line center. A one-size-fits-all approach to defining the continuum would easily clip out such broad tails, leading to systematic underestimates of the EWs. %Lastly, or 22 of them, we assigned EW = 0, whereas \citet{Fang2018} measured values anywhere between $-0.5$ and 0.3.

\subsubsection{Removing the \halpha\ Photospheric Contribution} \label{sec:ewcorr}

The observed EW values include both chromospheric emission, driven by magnetic heating, and the underlying photospheric absorption, whose depth depends primarily on stellar mass but may itself be modified by weak chromospheric heating \citep[e.g.,][]{Stauffer1986}. To account for this absorption, we estimated the depth of the quiescent (i.e., photospheric) absorption (EW$_\mathrm{quiescent}$) using the empirical relation of \citet[][hereafter N17]{Newton2017}, which predicts the maximum absorption envelope as a function of stellar mass{and is applicable to stars with masses $<$0.8 \Msun. As discussed by \citetalias{Newton2017}, weak chromospheric heating may initially strengthen \halpha\ absorption before increasing activity eventually fills the line and produces emission. Therefore, EW$_\mathrm{quiescent}$ should be regarded as an empirical reference level rather than a strictly photospheric, zero-activity contribution.

For stars with EW measurements and in the applicable stellar mass range, we subtracted EW$_\mathrm{quiescent}$ from the observed EW, yielding a relative EW that reflects only activity-related emission. Both the measured (i.e., observed) and relative values are reported in our catalog (see Table \ref{tbl:catalog}). Stars more massive than 0.8 \Msun\ are retained in our measured-EW comparisons, but all have \halpha\ firmly in absorption and therefore do not enter our subsequent chromospheric activity analysis in Section \ref{sec:ro_llh}.

\subsection{From \halpha\ Equivalent Widths to \llh} \label{sec:llh}

Whereas EW values provide a straightforward measure of \halpha\ activity, \llh\ is a more physically meaningful, mass-independent diagnostic. We derived \llh\ using the $\chi$ factor, the ratio of the continuum flux near \halpha\ to the bolometric flux \citep{Walkowicz2004}. This  allows for the conversion of an EW emission value into a luminosity ratio:
%\begin{equation}\label{eq:llh}
$L_{\mathrm{H}\alpha}/L_{\mathrm{bol}} = -\,EW \times \chi,$
%\end{equation}
where $EW$ is the absorption-corrected EW value. We calculated \llh\ only for stars with negative absorption-corrected EWs, corresponding to a measurable chromospheric \halpha\ emission component.

We calculated $\chi$ using the empirical \teff--$\chi$ relation of \citet{Nunez2024}, which is based on PHOENIX ACES model spectra \citep{Husser2013} and is an extension in \teff\ coverage of the relation in \citet{Douglas2014}. 
For each star, we linearly interpolated in the \teff--$\chi$ grid to determine $\chi$. We combined the resulting $\chi$ values, along with the relative (i.e., photosphere-corrected) \halpha\ EWs, to produce \llh. These values are included in our catalog (see Table \ref{tbl:catalog}) and form the basis for our analysis of chromospheric activity.

\section{Results and Discussion}\label{sec:results}

\begin{figure}
\centerline{\includegraphics{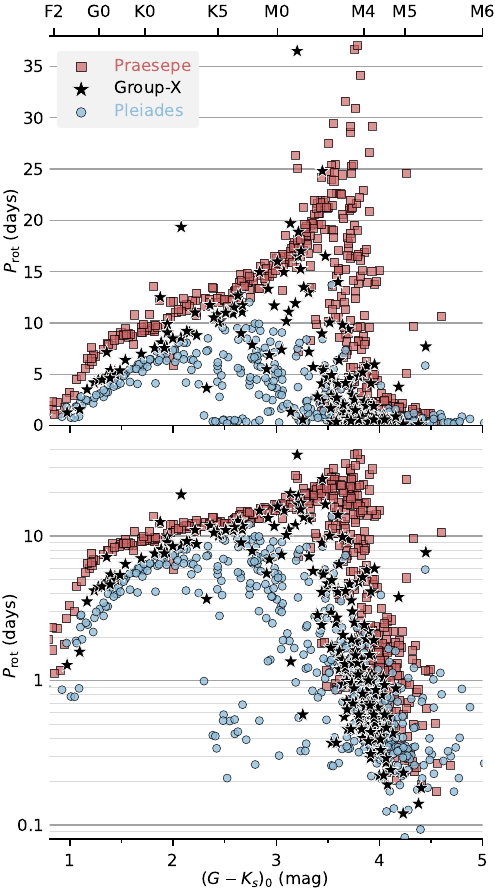}}
\vspace{-0.2cm}
\caption{Color vs.~\prot\ for single members of Group~X (black stars), Pleiades (blue circles), and Praesepe (red squares). The panels include the same data: the top panel shows \prot\ on a linear scale, and the bottom panel, on a log scale. The age of Group~X is between that of the Pleiades ($\approx$120 Myr old) and Praesepe ($\approx$700 Myr old), based on comparing the slow rotating sequence---from F2 to $\approx$K5 spectral types for Group~X---of the three clusters. While late K and early M dwarfs in Group~X have spun down relative to their counterparts in the Pleiades, they have similar periods to the late K and early M stars in Praesepe, implying that spin down for these stars is modified between the ages of Group~X and that of Praesepe.}
\label{fig:prot_clusters}
\end{figure}

\begin{figure*}
\centerline{\includegraphics{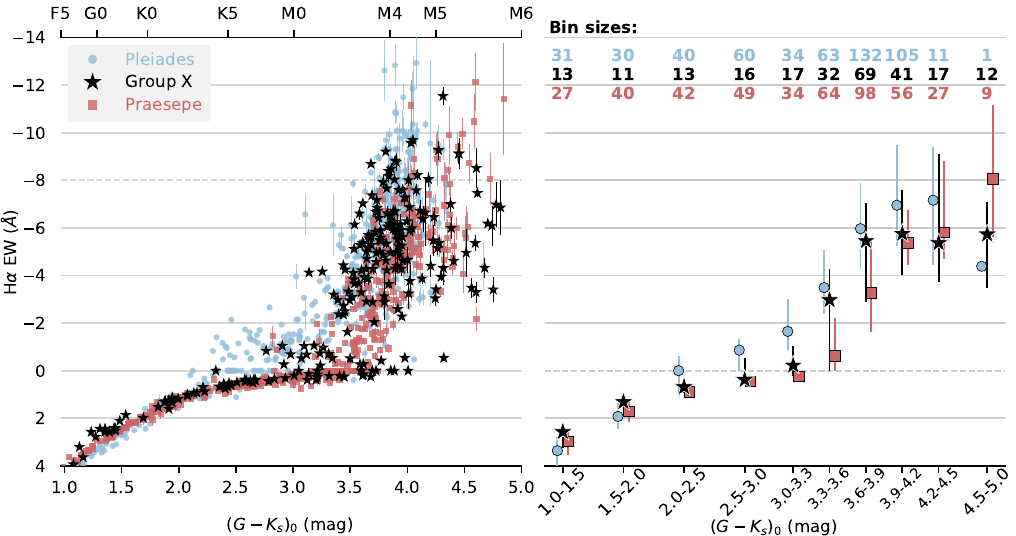}}
\vspace{0.0cm}
\caption{Color vs.~measured \halpha\ EW for single members of the Pleiades (blue circles), Group~X (black stars), and Praesepe \citep[red squares; data from][]{Nunez2024}. The left panel shows the full distribution of EW values for the F5--M6 spectral range. The right panel is the same data, binned by color (the bins are either 0.5 or 0.3 mag wide). The symbols represent the median of the measured \halpha\ EW values for each color bin, and the whiskers are the 16$^\mathrm{th}$ and 84$^\mathrm{th}$ percentiles. The numbers on top are the stars in each bin: the first row is for Pleiades, the second row is for Group~X, and the third row is for Praesepe.}
\label{fig:ew_clusters}
\end{figure*}

\subsection{Rotation Period Distribution and Gyrochronology}
Figure \ref{fig:prot_clusters} presents the distribution of stellar \prot\ as a function of color for single members of Group~X and the Pleiades, compared against an analogous data set for Praesepe ($\approx$700 Myr). For the latter, cluster membership and \prot\ data are from \citet{Nunez2024}; this rotation catalog includes measurements from K2 (the overwhelming majority of the \prot), TESS, and ZTF. %For the Pleiades, we consolidated several Gaia-based membership catalogs and used published rotational data from Kepler, K2, Palomar Transient Factory, and TESS (see Appendix \ref{app:pleiades}).

Binary systems are excluded, as their rotational evolution can be significantly altered by tidal effects or non-secular angular momentum loss \citep{Hurley2002, Fleming2019}. We used the binary flags in \citet{Nunez2024} to identify candidate binaries in Praesepe. We show the distribution of \prot\ values in both linear (top panel) and log (bottom) scales, to allow easier comparison of the slow-rotating sequences of the earlier spectral types (top panel) and of the full range of \prot\ for the later spectral types (bottom) for the three open clusters.

The slow-rotating sequence for Group~X, spans from the early F to the $\approx$K5 stars. In this spectral type range, Group~X stars are systematically located between the younger, faster rotating Pleiads and the older, slower rotating Praesepe stars. This  supports an age estimate for Group~X that is intermediate between $\approx$120 Myr and $\approx$700~Myr.

Beyond K5, interestingly, a number of slowly rotating members of Group~X have periods indistinguishable from the slow rotators of the same color in Praesepe, 
%In addition, some slowly rotating late K and early M dwarfs in Group X already overlap with the Praesepe slow-rotating sequence, 
despite the fact that the Praesepe stars are nearly twice as old. %$\approx$300 Myr age difference between the two clusters. 
This similarity in the \prot\ distributions is yet another illustration of stalled spin-down, which has been seen at older ages: \cite{Curtis2019}, building on work by \cite{Meibom2011}, showed that $\approx$1 Gyr-old K dwarfs in NGC~6811 have the same \prot\ as their younger counterparts in Praesepe, a behavior also seen for the handful of K dwarfs with measured \prot\ in the $\approx$1.4 Gyr-old cluster NGC 752 \citep{Agueros2018}.  \citet{Curtis2020} showed that these stalled stars eventually resume spinning down by $\approx$2.5 Gyr, the age of NGC 6819 and of Ruprecht 147, implying that rotation in these stars stalls for 1--2 Gyr. 

Our results for Group~X suggest that by 400 Myr, some of its late K and early M stars have entered a phase of reduced braking efficiency. The Group~X stars in this color range that are currently spinning more rapidly than their counterparts in Praesepe must join the slowly rotating sequence and stall by 700 Myr. The Group X data therefore push the evidence for the onset of stalled braking to younger ages and extend the evidence for this behavior into the late K/early M regime.

Metallicity differences between these clusters are unlikely to affect the interpretation of this comparison. Group X and the Pleiades have approximately solar metallicities \citep{Soderblom2009}, whereas Praesepe is modestly metal rich, with [Fe/H] $\approx +0.2$ \citep{Vejar2021, Agueros2025}. In principle, this could affect the color--\prot\ relation, because metallicity impacts stellar structure and the depth of the convective envelope, and may therefore alter magnetic braking efficiency \citep[e.g.,][]{Vansaders2013, Amard2020}. However, when \citet{Agueros2025} compared rotation distributions in the coeval Coma Berenices and Praesepe clusters, which also differ in metallicity by $\approx$0.2 dex, these authors found that the clusters' late F, G, K, and early M stars define remarkably similar color--\prot\ sequences. \citet{Agueros2025} concluded that any metallicity-driven differences in rotational behavior in these stars are negligible when working in the observed color--\prot\ plane (they do appear when working in the mass--\prot\ plane). The overlap between the Group X late K/early M slow rotators and the Praesepe sequence is therefore unlikely to be an artifact of Praesepe's enhanced metallicity, and more likely reflects genuinely reduced spin-down between the ages of Group X and Praesepe.

\subsection{The Evolution of Chromospheric Activity}\label{sec:chrom_dist}

Comparing \halpha\ EW distributions across clusters of different ages offers a direct view of the decay of chromospheric activity with time. As \halpha\ is a sensitive optical tracer of magnetic activity, in young clusters like the Pleiades many stars display strong \halpha\ emission tied to rapid rotation, while in older clusters such as Praesepe, spin-down leads to weaker emission. Positioning Group~X between these benchmarks shows how its members fit into this broader evolutionary sequence.

Figure \ref{fig:ew_clusters} shows the distribution of measured \halpha\ EW values as a function of color. We used the \halpha\ EW data in \citet{Nunez2024} for Praesepe, which were measured following the same procedures as in this study. We excluded cluster stars of spectral type earlier than F5 (\gminusk$_0\ \lapprox\ 0.97$ mag) from our analysis. Such stars lack a significant outer convective zone, and their magnetic activity is not driven by a solar-like dynamo, but likely by fossil magnetic fields \citep[e.g.,][]{Wolff1987, Boehm-Vitense2002, Kochukhov2003}.

The comparison with Pleiades and Praesepe stars again underscores the intermediate age of Group~X. In the left panel in Figure~\ref{fig:ew_clusters}, we see that for the Pleiades, the transition between \halpha\ absorption (positive EW values) and emission (negative EW values) occurs at \gminusk$_0$ $\approx$ 2.3 mag ($\approx$K4 spectral type). Group~X members transition from absorption to emission at \gminusk$_0$ $\approx$ 2.7 mag ($\approx$K7). In Praesepe, the transition occurs at \gminusk$_0$ $\approx$ 3.2 mag ($\approx$M1). This progression of the onset of \halpha\ emission to later spectral types from the Pleiades to Praesepe is consistent with the picture of decreasing stellar chromospheric activity with age.

In the right panel of Figure~\ref{fig:ew_clusters}, the color-binned version of the data allows us to quantify the difference in activity levels for different spectral types. The symbols represent the median values in each color bin and the whiskers are the 16$^\mathrm{th}$ and 84$^\mathrm{th}$ percentiles. We indicate the number of stars in each color bin at the top of the panel. For the one bin in the Pleiades with fewer than three stars, we only calculated the median EW. We defined smaller \gminusk$_0$ bin sizes in the 3.0--4.5 mag range to better sample the data distribution.

The binned data show that Pleiades stars across almost the full low-mass spectrum display stronger \halpha\ emission than their counterparts in the two older clusters, although the distributions show substantial overlap. For \gminusk$_0$ $\lapprox$ 2.0 mag (spectral types up to $\approx$K5), stars appear to converge to a tight \halpha\ absorption sequence in all three clusters. The transition from emission to absorption is also clear in the binned data: stars are near EW = 0 in the 2.0--2.5 mag color range for Pleiades, 2.5--3.0 mag range for Group X, and 3.0--3.3 mag range for Praesepe. The noticeable flattening of \halpha\ EW levels toward the red-most color bins (for \gminusk$_0$~$\gapprox$~3.9 mag) partly reflects the saturated level of magnetic activity for these late-type stars at the ages of the three clusters.

The comparison between rotation and \halpha\ activity near the K/M boundary reveals that the onset of stalled braking is not accompanied by a clear signal in the chromospheric data. Some late K/early M stars in Group X already lie on the weak-activity, photospheric \halpha\ absorption sequence traced by Praesepe, while others with similar colors and comparable \prot\ values still show measurable \halpha\ emission. Conversely, some faster-rotating Group X stars are already consistent with having minimal or undetectable \halpha\ emission at our spectral resolution. We illustrate this diversity in Figure \ref{fig:prot_ew}, where we color-code Group X stars (circles) by their measured EW values in the color-\prot\ plane against a background of Praesepe stars (white squares). This scatter suggests that during the transition into the stalled-braking phase, \prot\ alone does not determine the observed chromospheric output. %Instead, the \halpha\ activity level likely depends on additional factors, such as short-term variability, magnetic filling factor, magnetic topology, internal angular momentum evolution, or residual structural evolution near the ZAMS.

\begin{figure}
\centerline{\includegraphics{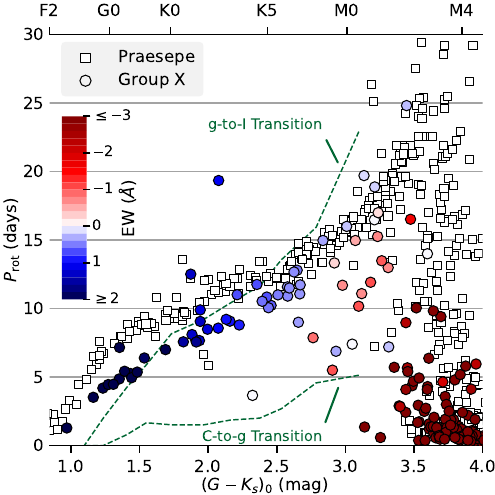}}
\vspace{-0.2cm}
\caption{Color vs.~\prot\ for single members of Group X (colored circles) and Praesepe (white squares). Group X stars are color-coded by their observed \halpha\ EW. The dashed green lines are the C-to-g and g-to-I transition relations from \citet{Yang2025}.}
\label{fig:prot_ew}
\end{figure}

This apparent mismatch between rotational and chromospheric evolution raises the question of whether the overlap between the Group X and Praesepe slow-rotator sequences necessarily indicates that these stars have reached the same evolutionary stage. \citet{Yang2025} recently proposed a framework that distinguishes between the visually converged slow-rotator sequence observed in young clusters and the more evolved interface (I) sequence. In their framework, stars evolve from the rapidly rotating convective (C) sequence through an intermediate gap phase (g) before reaching the zero-age I sequence (ZAIS), defined by the g-to-I transition. The authors associate the earlier, apparently converged population with a pre-I sequence and stalled spin-down. Their model therefore provides a possible interpretation of the rotational overlap we observe between Group X and Praesepe.

To explore this interpretation, we overplot the C-to-g and g-to-I transition relations from \citet{Yang2025} in Figure \ref{fig:prot_ew}, converting their \teff\ values to \gminusk$_0$ using the same empirical color--temperature relation adopted elsewhere in this work. For \teff\ bins representative of late-K and early-M stars, their model predicts C-to-g transition ages of $\approx$150--300 Myr and g-to-I transition ages of $\approx$0.8--1.0 Gyr. Many of the slowly rotating Group X stars in this color range lie below or near the predicted g-to-I boundary, consistent with their identification as a pre-I population that has not yet reached the ZAIS. Our \halpha\ measurements add another dimension to this picture: even among stars whose \prot\ appear to have converged, the observed chromospheric behavior remains diverse. Thus, although the \citet{Yang2025} framework offers a possible evolutionary interpretation of the rotational distributions, our results suggest that apparent rotational convergence need not be accompanied by uniform chromospheric activity.

\subsection{The Activity--Rotation Relation}\label{sec:ro_llh}
To examine the dependence of chromospheric activity on rotation quantitatively, we now consider \llh\ as a function of \Ro. Figure \ref{fig:rossby} shows the \Ro--\llh\ relation for single members of Group~X (top panel) and the Pleiades (bottom). 

\begin{figure}
\centerline{\includegraphics{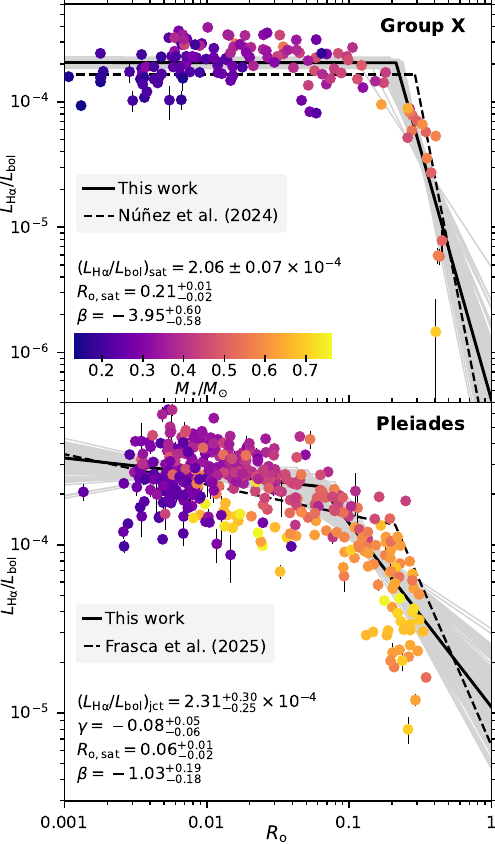}}
\vspace{-0.2cm}
\caption{\Ro\ vs.~\llh\ for single members of Group~X (top panel) and the Pleiades (bottom). Stars are color-coded by their $M_\star$. The solid black line is the maximum \textit{a posteriori} fit from the MCMC algorithms, and the gray lines are 200 random samples from the posterior probability distributions. For Group~X, we assumed a flat saturated regime described by (\llh)$_\mathrm{sat}$ and \Ro$_\mathrm{,sat}$, and an unsaturated regime described by a power-law of index $\beta$ (see Equation \ref{eq:rossby}). For the Pleiades, we replaced the flat saturated regime with one with a power-law of index $\gamma$, while maintaining continuity at the junction with the unsaturated regime (see Equation \ref{eq:rossby_sloped}). The annotated best fit results for each parameter are the median of each marginal posterior, with 16$^\mathrm{th}$--84$^\mathrm{th}$ percentiles as $\approx$1$\sigma$ intervals. Appendix \ref{app:corr} shows the marginalized posterior probability distributions from the MCMC analyses. The dashed lines indicate the results from similar analyses on a sample of Praesepe stars \citep[top panel;][]{Nunez2024} and Pleiads  \citep[bottom;][]{Frasca2025}.}
\label{fig:rossby}
\end{figure}

\subsubsection{Parametrization with a Flat Saturated Regime}
For Group~X, we modeled the relation as flat for \Ro\ $\leq$ \Ro$_\mathrm{,sat}$ with a power-law decline for \Ro\ $>$ \Ro$_\mathrm{,sat}$:
\begin{equation}\label{eq:rossby}
  \frac{L_{\mathrm{H}\alpha}}{L_{\mathrm{bol}}} = \left\{
  \begin{array}{l l}
    \left(\frac{L_{\mathrm{H}\alpha}}{L_{\mathrm{bol}}}\right)_{\mathrm{sat}} & \quad \textrm{if $R_\mathrm{o}\le R_{\mathrm{o,sat}}$}\\
    \left(\frac{L_{\mathrm{H}\alpha}}{L_{\mathrm{bol}}}\right)_{\mathrm{sat}}\left(\frac{R_\mathrm{o}}{R_{\mathrm{o,sat}}}\right)^{\beta} & \quad \textrm{if $R_\mathrm{o} > R_{\mathrm{o,sat}}$}
  \end{array} \right.
\end{equation}
where (\llh)$_\mathrm{sat}$ is the constant activity level in the saturated regime and $\beta$ is the slope of the unsaturated branch. This model has been commonly adopted in the literature \citep[e.g.,][]{Wright2011, Douglas2014, Newton2017}. 

We used the open-source Markov-chain Monte Carlo (MCMC) package \texttt{emcee} \citep{emcee} to fit this three-parameter model to our data in log--log space. The residuals about the best-fitting activity--rotation relation are approximately Gaussian with few outliers. We therefore adopted a Gaussian likelihood with an additive nuisance parameter $f$ to accommodate underestimated errors and intrinsic variability\footnote{See \url{https://emcee.readthedocs.io/en/develop/user/line/}.} \citep[e.g.,][]{Magaudda2020}. We assumed uniform priors within broad bounds over each parameter, with $\beta < 0$, and we used 300 walkers, each taking 5000 steps in their MCMC chain, to sample the posterior. We report the median of each marginal posterior as the point estimate with 16$^\mathrm{th}$--84$^\mathrm{th}$ percentiles as $\approx$1$\sigma$ intervals.

The parameters for the best fit to the Group~X data, shown in the top panel of Figure \ref{fig:rossby}, are (\llh)$_\mathrm{sat} = (2.06\pm0.07)\times 10^{-4}$, \Ro$_\mathrm{,sat}$ = $0.21^{+0.01}_{-0.02}$, and $\beta = -3.95^{+0.60}_{-0.58}$. We indicate the maximum \textit{a posteriori} model and 200 random samples from the posterior distributions as a solid black line and gray solid lines, respectively. The nuisance parameter $f$ converged to ln~$f = -3.2$, corresponding to $\approx$9.5\% fractional intrinsic scatter. Appendix \ref{app:corr} shows the MCMC posterior distributions for each parameter and 2D correlations between pairs of parameters from each fit.

In Figure \ref{fig:rossby}, we also include the results from our previous study for single members of Praesepe \citep[dashed line in top panel;][]{Nunez2024}. That study used the same techniques to obtain \llh\ from measured EWs and to calculate \Ro. Our best fit model very closely resembles that of Praesepe: the Group X (\llh)$_\mathrm{sat}$ is within 3$\sigma$ of the Praesepe value of (\llh)$_\mathrm{sat} = (1.76\pm0.09)\times10^{-4}$, the \Ro$_\mathrm{,sat}$ is also within 3$\sigma$ of \Ro$_\mathrm{,sat} = 0.28^{+0.02}_{-0.03}$ for the older cluster, and the $\beta$ is within 1$\sigma$ of $\beta = -5.19^{+1.32}_{-0.94}$ for Praesepe stars.

Similar to what was seen in \citet{Nunez2024}, the unsaturated regime is dominated by the more massive stars in our sample: the mean and 1$\sigma$ $M_\star$ for Group~X stars with \Ro\ $>$ \Ro$_\mathrm{,sat}$ is 0.56$\pm0.09$ \Msun, and for stars with \Ro\ $<$ \Ro$_\mathrm{,sat}$ is 0.32$\pm0.12$ \Msun. Evidently, as seen in the \prot\ distributions (Figure \ref{fig:prot_clusters}) and \halpha\ EW distributions (Figure \ref{fig:ew_clusters}), the decrease of rotation and chromospheric activity in coeval samples of stars occurs in an orderly manner, with more massive stars spinning down---and hence becoming less active---more quickly than less massive stars.

\subsubsection{Parametrization with a Sloped Saturated Regime}\label{sec:ro_llh_sloped}
Guided by the apparent slope in the saturated branch in the Pleiades reported by \citet{Fang2018} and \citet{Frasca2025}, we replaced the flat saturated regime in Equation \ref{eq:rossby} by a power-law decline that remains continuous at the junction with the unsaturated regime. This version of the model has also been implemented in different ways in the past \citep[e.g.,][]{Reiners2014, Magaudda2020, Han2024}.

The model described above then becomes:
\begin{equation}\label{eq:rossby_sloped}
  \frac{L_{\mathrm{H}\alpha}}{L_{\mathrm{bol}}} = \left\{
  \begin{array}{l l}
    \left(\frac{L_{\mathrm{H}\alpha}}{L_{\mathrm{bol}}}\right)_{\mathrm{jct}}\left(\frac{R_\mathrm{o}}{R_{\mathrm{o,sat}}}\right)^{\gamma} & \quad \textrm{if $R_\mathrm{o}\le R_{\mathrm{o,sat}}$}\\
    \left(\frac{L_{\mathrm{H}\alpha}}{L_{\mathrm{bol}}}\right)_{\mathrm{jct}}\left(\frac{R_\mathrm{o}}{R_{\mathrm{o,sat}}}\right)^{\beta} & \quad \textrm{if $R_\mathrm{o} > R_{\mathrm{o,sat}}$}
  \end{array} \right.
\end{equation}
where (\llh)$_\mathrm{jct}$ is the activity level at the junction \Ro\ = \Ro$_\mathrm{,sat}$ and $\gamma$ is the slope of the saturated branch. In the limit $\gamma \to 0$, the saturated branch becomes flat with \llh(\Ro $\le$ \Ro$_\mathrm{,sat}$) = (\llh)$_\mathrm{jct}$, and we recover Equation~\ref{eq:rossby}.

We implemented this model using the same procedures described above with small modifications. The Pleiades sample exhibits heavier-tailed residuals within the saturated branch, where flares and temporal variability are more common, and where the sloped-saturation fit gives extra leverage to low \Ro\ points. To obtain a fit that is robust to such outliers without \textit{ad hoc} clipping, we used a Student-$t$ likelihood with $\nu = 5$ degrees of freedom. We also limited $|\gamma|$ to modest values. 

The best-fit parameter values for the Pleiades are (\llh)$_\mathrm{jct} = (2.31^{+0.30}_{-0.25})\times 10^{-4}$, $\gamma = -0.08^{+0.05}_{-0.06}$, \Ro$_\mathrm{,sat}$ = $0.06^{+0.01}_{-0.02}$, and $\beta = -1.03^{+0.19}_{-0.18}$. We also calculated \llh\ at the median \Ro\ for \Ro\ $<$ \Ro$_\mathrm{,sat}$, as a reference activity level in the saturated regime. We computed this (\llh)$_\mathrm{sat,ref}$ value for every posterior draw, and the median value (and 16$^\mathrm{th}-84^\mathrm{th}$ percentiles) of the distribution is $(2.66\pm0.09)\times 10^{-4}$. Lastly, $f$ converged to ln~$f = 0.15$, which corresponds to $\approx$35\% fractional intrinsic scatter. 

The bottom panel of Figure \ref{fig:rossby} shows our results for Pleiades single stars. The fitted slope in the saturated regime ($\gamma = -0.08^{+0.05}_{-0.06}$) reflects the slight---albeit barely statistically significant---dependence of chromospheric activity on rotation among the fastest rotators. To place this in context, we compared our best-fit parameters with those reported by \citet{Frasca2025} and \citet{Fang2018}, whose \halpha\ measurements were also obtained from LAMOST spectra---the latter study, however, using medium-resolution spectra. Both studies identified a negative slope within the saturated regime, of order $-0.2$, within 3$\sigma$ of our result. However, the methodological details differ substantially. \citet{Frasca2025} explicitly adopted \Ro$_\mathrm{,sat} = 0.21$ from \citetalias{Newton2017} rather than fitting for it, while \citet{Fang2018} simply estimated \Ro$_\mathrm{,sat} \approx 0.2$ and the slope in the unsaturated regime by eye.

Our MCMC approach refines these earlier analyses by jointly constraining the transition point between the two regimes, \Ro$_\mathrm{,sat}$, and the power-law slopes while accounting for intrinsic scatter and heavy-tailed residuals. The resulting parameters reproduce the same qualitative behavior, but provide a more statistically grounded quantification of the Pleiades activity--rotation relation.

The persistence of a negative slope in the saturated branch indicates that, at $\approx$120 Myr, the most rapidly rotating stars have not yet reached a steady saturation state. Rather than reflecting an increase in magnetic heating efficiency with increasing rotation, the negative slope in the saturated regime may arise from the combined effects of short-term variability and flare-enhanced emission \citep[e.g.,][]{Magaudda2020,Raetz2020}, and/or from modest structural and dynamo-regime differences that modulate chromospheric output at the highest activity levels \citep[e.g.,][]{Donati2011, Gregory2012, Reiners2014, Gully-Santiago2017}.

By contrast, the Group~X sample shows no measurable slope in its saturated branch. We tested fitting the model in Equation \ref{eq:rossby_sloped} to our Group~X sample but found $\gamma$ to be statistically indistinguishable from zero. The flattening of the relation between the Pleiades and Group~X may indicate the emergence of a stable saturation regime, potentially linked to changes in dynamo operation, magnetic topology, or the declining contribution of pre-main-sequence stars in the saturated regime.

A further clue may come from the structural evolution of the stars populating the saturated branch at these ages. At the Pleiades age ($\approx$120 Myr), stars with $M_\star\ \lapprox\ 0.7$ \Msun---essentially all late K and M dwarfs, which dominate the small \Ro\ regime---are still on their pre-MS contraction tracks and remain well above the ZAMS \citep{Baraffe2015, Somers2017}. Their luminosities and radii continue to decline as gravitational contraction releases additional energy, potentially fueling enhanced magnetic heating that is not yet regulated by a steady, rotation-driven dynamo, as mentioned before. In such a population, the saturated branch likely does not represent a single, time-independent saturation limit; it may instead combine stars at slightly different structural stages, with different flare rates and quiescent chromospheric emission levels. By the age of Group~X ($\approx$400 Myr), only stars with $M_\star\ \lapprox\ 0.5$ \Msun\ (M1 and later) remain pre-ZAMS, and thus a significant fraction of the saturated population has reached the MS. The disappearance of these structurally evolving stars from the saturated population may contribute to the emergence of the flat saturation plateau seen in Group X.

\subsection{The Activity--Rotation Relation Over Time}\label{sec:ro_time}

We compared the \Ro--\llh\ distributions of single stars of the Pleiades, Group~X, and Praesepe and of the large sample of nearby M dwarfs in \citetalias{Newton2017}. This allowed us to trace how stars of different spectral types evolve from the saturated to the unsaturated regime as they spin down with age.

\subsubsection{Revisiting the M Dwarf Sample in \citet{Newton2017}}
To obtain the cleanest single-star field-age sample for this comparison, we first searched the \citetalias{Newton2017} catalog for any unidentified candidate binaries. We found Gaia DR3 counterparts to the \citetalias{Newton2017} stars and flagged as candidate binaries those with RUWE $>$ 1.4 or those present in Gaia's NSS catalog.

Additionally, we removed from the \citetalias{Newton2017} catalog stars that are likely members of nearby young associations. For this, we used the multivariate Bayesian algorithm \texttt{BANYAN $\Sigma$} \citep{Gagne2018}, which can identify in 6-dimensional space if a star is more likely to be part of one of 27 known young associations than to be a field-age star.\footnote{The 27 known associations included in \texttt{BANYAN $\Sigma$} span the age range 1--800 Myr and are within 150 pc of the Sun. We replaced the \citet{Perryman1998} Hyades ($\approx$700 Myr old) catalog in the \texttt{BANYAN $\Sigma$} list of reference catalogs with the more recent \citet{Nunez2022} catalog.} As input values for \texttt{BANYAN $\Sigma$}, we used Gaia DR3 parallax, proper motion, and radial velocity data, occasionally using values from SIMBAD when Gaia data were incomplete.

We also updated \Ro\ and \llh\ values for stars in our cleaned sample. We first derived their $M_\star$ and \teff\ using 2MASS $K_s$ magnitudes and following the steps described in Section \ref{sec:properties}. Next, we calculated $\tau$ using the updated $M_\star$--$\tau$ relation in \citet{Wright2018} to obtain new \Ro\ values. Lastly, we calculated $\chi$ using the extended \teff--$\chi$ relation in \citet{Nunez2024} and the quiescent \halpha\ EW with the $M_\star$--EW$_\mathrm{quiescent}$ relation of \citetalias{Newton2017} (see Section \ref{sec:ewcorr}) to obtain updated \llh\ values.

On average, our \llh\ values are $\approx$20\% higher than those of \citetalias{Newton2017} and our \Ro\ values are $\approx$20\% lower. These disparities reflect the slight differences in the calculation of fundamental stellar properties. Using our updated values, we redid the best fit to the \Ro--\llh\ distribution of the cleaned \citetalias{Newton2017} M dwarf sample, as described in Section \ref{sec:ro_llh}---a procedure that closely resembles the fitting performed in \citetalias{Newton2017}. Our best fit parameters \llh, \Ro$_\mathrm{sat}$, and $\beta$ are all within one sigma of those in \citetalias{Newton2017}. Evidently, the impact of updating \llh\ and \Ro\ values while simultaneously excluding potential binaries and young stars  results in a statistically unchanged fit to the \Ro--\llh\ distribution.

\begin{figure*}
\centerline{\includegraphics{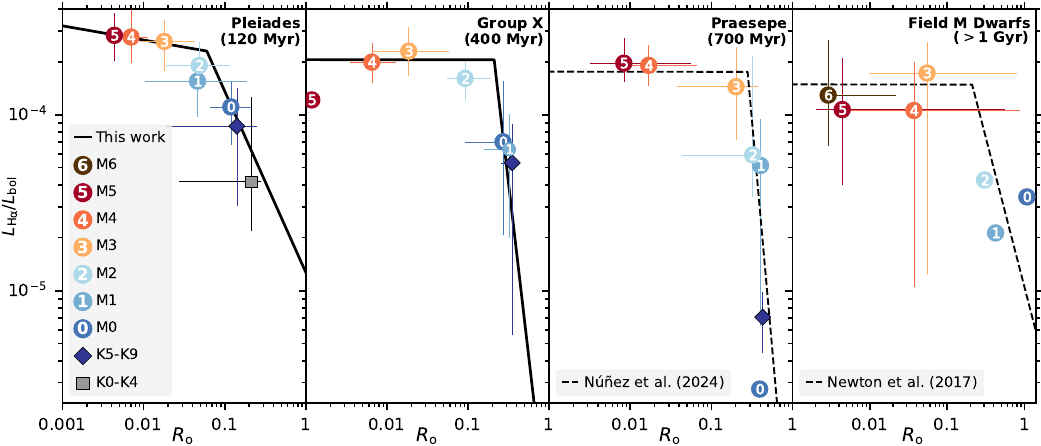}}
\vspace{-0.2cm}
\caption{\Ro\ vs.~\llh\ binned by spectral type for single members of the Pleiades (leftmost panel), Group~X (center left), Praesepe (center right), and our cleaned field age M dwarf sample from the full M dwarf sample in \citetalias{Newton2017} (rightmost panel). The symbols represent the median of each bin, and the whiskers are the 16$^\mathrm{th}$ and 84$^\mathrm{th}$ percentiles. Symbols without whiskers are bins with fewer than three stars. The solid line in the two left-most panels is our maximum \textit{a posteriori} fit for each sample (see Figure~\ref{fig:rossby}). The dashed lines in the two right-most panels are the maximum \textit{a posteriori} fits for Praesepe from \citet{Nunez2024} and for the full M dwarf sample in \citetalias{Newton2017}. The bin statistics presented here are listed in Table \ref{tbl:rossby_stats}.}
\label{fig:rossby_binned}
\end{figure*}

\subsubsection{The \Ro--\llh\ Relation Across Time}
Figure \ref{fig:rossby_binned} presents the median \llh\ and \Ro\ values (with 16$^\mathrm{th}$--84$^\mathrm{th}$ percentile ranges) for spectral type bins for the Pleiades, Group~X, Praesepe, and the cleaned field-age star sample of nearby M dwarfs from \citetalias{Newton2017} (presumed ages $\gapprox$ 1 Gyr). We converted \gminusk$_0$ into spectral types for all four samples using the empirical relations tabulated by E. Mamajek and included in Table \ref{tbl:catalog}.\footnote{See Footnote \ref{note}.} Table \ref{tbl:rossby_stats} includes all the \llh\ and \Ro\ median and 16$^\mathrm{th}$--84$^\mathrm{th}$ percentiles for the four samples. The solid lines in the Pleiades and Group~X panels are our best fits from this work, and the dashed lines in the Praesepe and field-age M dwarfs panels are the best fit from \citet{Nunez2024} and \citetalias{Newton2017}, respectively.

A clear evolutionary pattern emerges: the higher mass K dwarfs (represented in Figure \ref{fig:rossby_binned} by squares and diamonds) are the first to transition out of the saturated regime, as seen by the age of the Pleiades, while progressively later spectral types move toward the unsaturated branch as the clusters age. Among M dwarfs (represented in Figure \ref{fig:rossby_binned} by colored circles, with the number in each circle indicating the corresponding M spectral subtype), this progression unfolds sequentially with subtype: M0 and M1 stars approach the unsaturated regime by the age of the Pleiades and Group X, followed by M2 stars in Praesepe, with still later types remaining largely saturated at these ages. The same general trend of decreasing activity with increasing \Ro\ is discernible even in the field-age sample (right-most panel in Figure \ref{fig:rossby_binned}). The comparatively larger \llh\ 16$^\mathrm{th}$--84$^\mathrm{th}$ percentile ranges for the latter sample reflect the inherent heterogeneity of the stars included, which can range in age from $\approx$1 Gyr to potentially the age of the Galaxy.

Our results suggest that M3 and later stars remain saturated well past ages of several 100 Myr. However, their \llh\ levels show a modest but systematic decline with age, on the order of $\approx$0.1--0.3 dex between the $\approx$120 Myr-old sample and the field M dwarf sample. Indeed, (\llh)$_\mathrm{sat}$ for Group X stars is twice that for the sample in \citetalias{Newton2017}.

This decline in saturated activity level may reflect a combination of effects. First, the frequency or energetic output of flares may decrease with age, reducing the time-averaged chromospheric emission of young M dwarfs \citep[e.g.,][]{Loyd2018, Raetz2020}. Second, the quiescent chromospheric emission level itself may decline even before stars formally leave the saturated regime. The pre-MS context introduced above provides a natural route for such an effect: at $\approx$120 Myr, the M dwarfs that dominate the saturated population are still pre-MS and may exhibit enhanced magnetic emission, whereas by $\approx$400 Myr only M3 and later types remain pre-MS, and by $\approx$700 Myr this fraction is limited to M4 types and later \citep{Tognelli2011, Baraffe2015}. The progressive loss of these structurally evolving, magnetically active pre-MS members would shift the mean saturated activity level downward with age, even if the underlying dynamo saturation limit were unchanged. Testing this interpretation will require comparing chromospheric activity for stars spanning the full low-mass range before and after ZAMS arrival, to isolate structural evolution from the secular decay of magnetic activity.

\begin{deluxetable}{@{}lcccc@{}}
\tabletypesize{\footnotesize}

\tablecaption{Activity--Rotation Relation Statistics \label{tbl:rossby_stats}}

\tablehead{
\colhead{} & \colhead{Pleiades} & \colhead{Group X} & \colhead{Praesepe} & \colhead{Field Age}\\[-0.09 in]
 &  &  & & \colhead{M Dwarfs}\\[-0.2 in]
}
\startdata
\multicolumn{5}{c}{\it \small Number of Stars Per Bin} \\[0.0 in]
K0--K4 & 11 &  0 &  0 &  0\\
K5--K9 & 66 &  7 & 15 &  0\\
M0    &  16 &  4 &  1 &  1\\
M1    &  11 &  6 &  7 &  1\\
M2    &  26 & 12 & 15 &  2\\
M3    & 116 & 57 & 83 & 27\\
M4    & 151 & 49 & 64 & 68\\
M5    &   8 &  2 & 12 & 24\\
M6    &   0 &  0 &  0 & 4\\
\tableline
Total & 396 & 138 & 197 & 127\\[0.05 in]
\multicolumn{5}{c}{\it \small Median \Ro} \\[0.02 in]
K0--K4 & 0.213$^{+0.068}_{-0.186}$ & \nodata                   & \nodata                   & \nodata \\
K5--K9 & 0.142$^{+0.112}_{-0.123}$ & 0.352$^{+0.055}_{-0.098}$ & 0.430$^{+0.021}_{-0.020}$ & \nodata \\
M0     & 0.119$^{+0.092}_{-0.053}$ & 0.270$^{+0.085}_{-0.178}$ & 0.395                     & 1.074\\
M1     & 0.046$^{+0.140}_{-0.036}$ & 0.321$^{+0.066}_{-0.164}$ & 0.404$^{+0.026}_{-0.050}$ & 0.423\\
M2     & 0.049$^{+0.069}_{-0.030}$ & 0.092$^{+0.100}_{-0.036}$ & 0.323$^{+0.081}_{-0.279}$ & 0.305\\
M3     & 0.018$^{+0.025}_{-0.009}$ & 0.019$^{+0.040}_{-0.012}$ & 0.201$^{+0.170}_{-0.163}$ & 0.055$^{+0.754}_{-0.045}$\\
M4     & 0.007$^{+0.004}_{-0.002}$ & 0.007$^{+0.006}_{-0.003}$ & 0.017$^{+0.049}_{-0.010}$ & 0.037$^{+0.841}_{-0.031}$\\
M5     & 0.004$^{+0.001}_{-0.001}$ & 0.001                     & 0.008$^{+0.047}_{-0.005}$ & 0.004$^{+0.550}_{-0.002}$\\
M6     & \nodata                   & \nodata                   & \nodata                   & 0.003$^{+0.019}_{-0.001}$\\[0.05 in]
\multicolumn{5}{c}{\it \small Median \llh\ (10$^{\it -4}$)} \\[0.02 in]
K0--K4 & 0.42$^{+0.84}_{-0.20}$ & \nodata                & \nodata                & \nodata\\
K5--K9 & 0.86$^{+0.56}_{-0.56}$ & 0.53$^{+0.36}_{-0.48}$ & 0.07$^{+0.03}_{-0.03}$ & \nodata\\
M0     & 1.11$^{+0.43}_{-0.43}$ & 0.70$^{+0.85}_{-0.50}$ & 0.03                   & 0.34\\
M1     & 1.55$^{+0.57}_{-0.58}$ & 0.64$^{+0.38}_{-0.44}$ & 0.52$^{+0.43}_{-0.45}$ & 0.21\\
M2     & 1.91$^{+0.70}_{-0.46}$ & 1.61$^{+0.42}_{-0.39}$ & 0.59$^{+1.57}_{-0.25}$ & 0.43\\
M3     & 2.61$^{+0.83}_{-0.83}$ & 2.30$^{+0.86}_{-0.64}$ & 1.45$^{+0.98}_{-0.73}$ & 1.73$^{+0.86}_{-1.60}$\\
M4     & 2.77$^{+1.24}_{-0.79}$ & 1.99$^{+0.58}_{-0.48}$ & 1.91$^{+0.59}_{-0.46}$ & 1.06$\pm0.95$\\
M5     & 2.83$^{+0.94}_{-0.81}$ & 1.22                   & 1.97$^{+0.76}_{-0.44}$ & 1.07$^{+1.04}_{-0.67}$\\
M6     & \nodata                & \nodata                & \nodata                & 1.30$^{+1.35}_{-0.63}$\\[0.02 in]
\enddata

\tablecomments{The quoted uncertainties correspond to the 16$^\mathrm{th}$ and 84$^\mathrm{th}$ percentiles. No uncertainties are quoted for bins with fewer than three stars.}

\end{deluxetable}

\section{Conclusion}\label{sec:conclusion}

We carried out the first large and uniform analysis of chromospheric activity in the $\approx$400 Myr-old open cluster Group X. We constructed a consolidated Gaia-based membership catalog by merging several recent literature studies, producing a robust sample of single low-mass stars suitable for comparing rotation and chromospheric activity across clusters. To characterize Group X, we combined new optical spectroscopy from the MDM 2.4 m Hiltner telescope with archival LAMOST and SDSS spectra, yielding low-resolution spectra for 323 members. From these data we measured \halpha\ EWs for nearly two-thirds of the low-mass population, and derived \llh\ using updated $\chi$ factors and empirically calibrated photospheric corrections. We paired these measurements with literature \prot\ data drawn from TESS and ZTF surveys. We expanded the Group X \prot\ coverage using more recent TESS and ZTF light curves.

We performed a parallel effort for the Pleiades, assembling a comprehensive Gaia-based membership catalog from multiple recent studies and compiling \prot\ data from HATNet, K2, PTF, TESS, and TARS from the literature. For chromospheric activity, we used LAMOST low-resolution spectra for 612 members to obtain \halpha\ measurements, ensuring consistent \llh\ measurements for all clusters. These consolidated samples allowed a direct comparison of Group X to the younger Pleiades and the older ($\approx$700 Myr old) Praesepe cluster.

The \prot\ distributions show that Group X falls between the Pleiades and Praesepe for F through mid K stars, but late K and early M dwarfs in Group X already overlap the Praesepe slow-rotator sequence. This suggests that the stalled spin-down phase previously identified in $\approx$1 Gyr-old clusters is already present by $\approx$400 Myr for stars near the K/M boundary. Group X therefore constrains the onset of stalled braking, complementing older clusters such as NGC 6811, NGC 752, and Ruprecht 147.

The \halpha\ EW distributions reinforce this evolutionary picture. The spectral type at which stars transition from \halpha\ absorption to emission shifts steadily toward lower masses with age: from roughly mid K in the Pleiades, to late K in Group X, and to early M in Praesepe. Group X therefore marks the stage at which chromospheric activity has already weakened substantially among K dwarfs, while remaining strong across much of the M dwarf population. This behavior provides the empirical bridge between the highly active low-mass stars in the Pleiades and the less active population in Praesepe.

The comparison between these two diagnostics shows that rotational and chromospheric evolution are not fully synchronized during this epoch. Although late K and early M stars in Group X already overlap the Praesepe slow-rotator sequence, their \halpha\ EWs do not collapse onto a single activity sequence: some resemble weakly active Praesepe stars, while others retain stronger emission. The rotational behavior of these stars is qualitatively consistent with the pre-I sequence proposed by \citet{Yang2025}, in which stars can exhibit apparent rotational convergence before reaching the mature I sequence. Our results add a chromospheric dimension to this picture, suggesting that apparent rotational convergence need not be accompanied by uniform \halpha\ activity.

Also revealing is how Group X reshapes the activity--rotation relation compared to the Pleiades. The former cluster exhibits a genuinely flat saturated regime in \llh, in contrast to the latter, whose saturated branch displays a shallow power law decline with increasing \Ro. This shift between $\approx$120 and $\approx$400 Myr marks the emergence of a stable saturation plateau: at Pleiades age, pre-MS stars dominate the saturated regime, and enhanced levels of saturated chromospheric activity in such stars may trace a transient phase where magnetic energy generation is still coupled to contraction, not solely to rotation.

Examined across spectral types, our results highlight a mass-dependent progression through the saturated and unsaturated regimes: K and early-M dwarfs begin to desaturate by Pleiades age; their activity levels decrease gradually across the ages of Group X and Praesepe; and mid-to-late M dwarfs remain saturated through several hundred Myr---and even field ages---but already display a measurable decline in their saturated activity levels. This decreasing saturated level of \llh\ likely reflects both rotational spin-down and the diminishing contribution of pre-MS stars---whose inflated radii and elevated magnetic output shape the youngest populations.

\begin{acknowledgments}
We thank Justin Anderson, Sofia Lawsky, Justin Rupert, and Luna Wang for their assistance in obtaining spectra for this work at the MDM Observatory.

A.N.~acknowledges support provided by the NSF through grant 2138089. M.A.A.and J.L.C.~acknowledge support provided by NASA through grant 80NSSC21K0989. M.A.A.~acknowledges additional support provided by NASA through TESS grant 80NSSC23K0370 (program ID G05028).

This work is based on observations obtained at the MDM Observatory, operated by Dartmouth College, Columbia University, Ohio State University, Ohio University, and the University of Michigan. The authors are honored to be permitted to conduct astronomical research on Iolkam Du'ag (Kitt Peak), a mountain with particular significance to the Tohono O'odham.

This paper includes data collected by the TESS mission, which are publicly available from the Mikulski Archive for Space Telescopes (MAST). Funding for the TESS mission is provided by NASA’s Science Mission directorate. The TESS data used in this paper can be found in MAST: \dataset[10.17909/0cp4-2j79]{http://dx.doi.org/10.17909/0cp4-2j79}.

This work is based on observations obtained with the Samuel Oschin Telescope 48-inch and the 60-inch Telescope at the Palomar Observatory as part of the Zwicky Transient Facility project \citep{ZTF}. ZTF is supported by the National Science Foundation under Grants No. AST-1440341 and AST-2034437 and a collaboration including current partners Caltech, IPAC, the Weizmann Institute for Science, the Oskar Klein Center at Stockholm University, the University of Maryland, Deutsches Elektronen-Synchrotron and Humboldt University, the TANGO Consortium of Taiwan, the University of Wisconsin at Milwaukee, Trinity College Dublin, Lawrence Livermore National Laboratories, IN2P3, University of Warwick, Ruhr University Bochum, Northwestern University and former partners the University of Washington, Los Alamos National Laboratories, and Lawrence Berkeley National Laboratories. Operations are conducted by COO, IPAC, and UW. ZTF data used in this paper can be found in IPAC: \dataset[10.26131/IRSA598]{https://doi.org/10.26131/IRSA598}.

Funding for the Sloan Digital Sky Survey V has been provided by the Alfred P. Sloan Foundation, the Heising-Simons Foundation, the National Science Foundation, and the Participating Institutions. SDSS acknowledges support and resources from the Center for High-Performance Computing at the University of Utah. SDSS telescopes are located at Apache Point Observatory, funded by the Astrophysical Research Consortium and operated by New Mexico State University, and at Las Campanas Observatory, operated by the Carnegie Institution for Science. The SDSS website is \url{www.sdss.org}.

SDSS is managed by the Astrophysical Research Consortium for the Participating Institutions of the SDSS Collaboration, including Caltech, The Carnegie Institution for Science, Chilean National Time Allocation Committee (CNTAC) ratified researchers, The Flatiron Institute, the Gotham Participation Group, Harvard University, Heidelberg University, The Johns Hopkins University, L'Ecole polytechnique f\'{e}d\'{e}rale de Lausanne (EPFL), Leibniz-Institut f\"{u}r Astrophysik Potsdam (AIP), Max-Planck-Institut f\"{u}r Astronomie (MPIA Heidelberg), Max-Planck-Institut f\"{u}r Extraterrestrische Physik (MPE), Nanjing University, National Astronomical Observatories of China (NAOC), New Mexico State University, The Ohio State University, Pennsylvania State University, Smithsonian Astrophysical Observatory, Space Telescope Science Institute (STScI), the Stellar Astrophysics Participation Group, Universidad Nacional Aut\'{o}noma de M\'{e}xico, University of Arizona, University of Colorado Boulder, University of Illinois at Urbana-Champaign, University of Toronto, University of Utah, University of Virginia, Yale University, and Yunnan University.

Guoshoujing Telescope (the Large Sky Area Multi-Object Fiber Spectroscopic Telescope LAMOST) is a National Major Scientific Project built by the Chinese Academy of Sciences. Funding for the project has been provided by the National Development and Reform Commission. LAMOST is operated and managed by the National Astronomical Observatories, Chinese Academy of Sciences.

This work has made use of data from the European Space Agency (ESA) mission Gaia (\url{https://www.cosmos.esa.int/gaia}), processed by the Gaia Data Processing and Analysis Consortium (DPAC, \url{https://www.cosmos.esa.int/web/gaia/dpac/consortium}). Funding for the DPAC has been provided by national institutions, in particular the institutions participating in the Gaia Multilateral Agreement. Gaia data can be found in IPAC: \dataset[10.26131/IRSA544]{https://doi.org/10.26131/IRSA544}.

This publication makes use of data products from the Two Micron All Sky Survey, which is a joint project of the University of Massachusetts and the Infrared Processing and Analysis Center/California Institute of Technology, funded by the National Aeronautics and Space Administration and the National Science Foundation. The data used in this paper can be found in IPAC: \dataset[10.26131/IRSA2]{https://catcopy.ipac.caltech.edu/dois/doi.php?id=10.26131/IRSA2}.

This research made use of the SIMBAD database and the VizieR catalog access tool, CDS, Strasbourg Astronomical Observatory, France. It also made use of ChatGPT (OpenAI 2026) for assistance in correcting typos and grammar, though full responsibility for the manuscript's content remains our own.

\end{acknowledgments}

%\begin{contribution}
%%This section gives authors the space to recognize author contributions. The text inside this environment is NOT counted towards the total word quanta. At a minimum, manuscripts are expected to include this text:

%All authors contributed equally to this work.

%% But authors are expected to provide more specific details, e.g. 
%%
%%SC was responsible for writing and submitting the manuscript.
%%WWM came up with the initial research concept and edited the manuscript.
%%OTS obtained the funding and edited the manuscript.
%%EBF provided the formal analysis and validation. He also edited the manuscript.
%%GEH Supervised the undergraduates, wrote the software and administers the project github and Zenodo repositories.
%%
%% Authors can use the Contributor Role Taxonomy (CRediT) at
%% for ideas on how write a good statement tailored to their needs.
%\end{contribution}

\facilities{Gaia, Hiltner(OSMOS), LAMOST, PO:1.2m(ZTF), SIMBAD, Sloan, TESS, VizieR}

%% Authors should list each code and include either a citation or url to the code inside ()s when available.
\software{astropy \citep{astropyIII},
          astroquery \citep{astroquery},
          BANYAN $\Sigma$ \citep{Gagne2018},
          corner \citep{corner}
          emcee \citep{emcee},
          IDL Astronomy User's Library \citep{IDLastro},
          Matplotlib \citep{matplotlib},
          NumPy \citep{numpy},
          PHEW \citep{phew},
          PypeIt \citep{pypeit, pypeit_zenodo},
          PySpecKit \citep{pyspeckit},
          SciPy \citep{scipy},
          TESScut, 
          unpopular}

\bibliography{main}{}
\bibliographystyle{aasjournalv7}



\appendix

\section{Selecting \prot\ Values for Certain Pleiads}\label{app:prots}
There are 39 Pleiads with \prot\ values from more than one of the studies mentioned in Section \ref{sec:prots_pl} that are discrepant by $>$15\%. We describe next how we selected the most appropriate \prot\ value for each of these 39 cases:

\begin{itemize}
\item \object{Gaia DR3 65214409762926720}: The star's $v \sin{i}$ in \citet{Queloz1998} agrees with the TESS-derived \prot\ = 7.04 days but not with the HATNet-derived \prot\ = 0.90 days; we adopted the former.
\item \object{Gaia DR3 65241313437941504}: The K2-derived \prot\ = 17.28 days is unusually long for the star's color, \gminusk$_0$ = 4.1 mag ($\approx$M4), and its age; we adopted instead the HATNet-derived \prot\ = 0.94 days.
\item \object{Gaia DR3 67920926352940032}: There is an eclipsing binary $\approx$6\arcmin\ away with a 0.37 day orbital period \citep{Drake2014}; we suspect the eclipsing binary signal is blended in the TESS light curve used in the TARS. Clipping out the transits, we confirmed the PTF-derived value (2.76 days) over the TARS-derived value (1.38 days or half that found with PTF). 
\item \object{Gaia DR3 68580530251050496}: The PTF and HATNet \prot\ values disagree by $\approx$30\% (0.30 versus 0.42 days, respectively); we adopted the latter, as HATNet had a shorter observing cadence than PTF (both studies had equivalent time baselines).
\item \object{Gaia DR3 69863454162150656}: We adopted the TARS \prot\ (0.688 days) over the HATNet \prot\ (0.407 days) because we did not recover the latter signal in the TESS light curves. %; we assumed that the higher TESS cadence enabled a more accurate period measurement.
\item There are seven stars with three \prot\ values where one is  discrepant from the other two by $>$15\%; in these cases, we ignored the discrepant \prot\ and adopted the mean of the remaining two.
\item For three stars with discrepant PTF and TARS \prot\ values, we adopted the TARS \prot, all of which are below 2.8 days, as we have seen PTF/ZTF sometimes struggle to correctly identify fast rotators due to the PTF/ZTF cadence.
\item For two stars with discrepant HATNet and TARS \prot\ values, we adopted the latter. The TARS \prot\ are precisely twice the HATNet values, and we visually confirmed these periods with the TESS light curves. 

\end{itemize}

In the remaining 22 cases, the decision was easier: we adopted the K2-derived \prot\ by virtue of K2's lack of atmospheric interference, smaller pixel size, and in most cases longer light curve duration.

\section{Marginalized Posterior Probability Distributions for the MCMC Analysis of Our \Ro--\llh\ Models}\label{app:corr}

We present the marginalized posterior probability distributions from the MCMC analysis we performed on our sample of \textit{bona fide} single members of Group~X (Equation \ref{eq:rossby}) and the Pleiades (Equation \ref{eq:rossby_sloped}; see Section \ref{sec:ro_llh}). We used the open-source package \texttt{corner} \citep{corner} to visualize these distributions.

\begin{figure*}
\centerline{\includegraphics{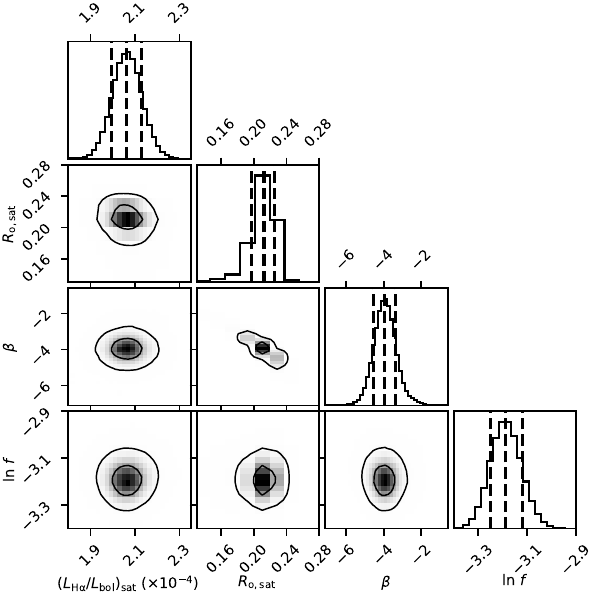}}
\vspace{-0.2cm}
\caption{Marginalized posterior probability distributions from the MCMC analysis of our \Ro--\llh\ model with a flat saturated regime using \texttt{emcee} for single members in Group~X. The parameter values of the \textit{a posteriori} model are the peaks of the one-dimensional distributions; the vertical dashed lines approximate the 16$^\mathrm{th}$, 50$^\mathrm{th}$, and 84$^\mathrm{th}$ percentiles. The two-dimensional distributions illustrate covariances between parameters; the contour lines approximate the 1$\sigma$ and 2$\sigma$ levels of the distributions.}
\label{fig:corr}
\end{figure*}

\begin{figure*}
\centerline{\includegraphics{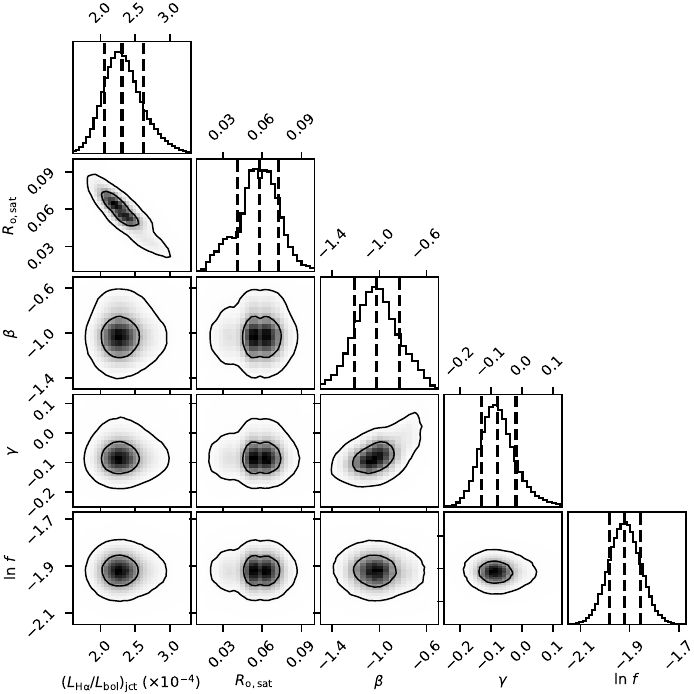}}
\vspace{-0.2cm}
\caption{Same as Figure \ref{fig:corr}, but for the \Ro--\llh\ model with a power-law component in the saturated regime for single members of the Pleiades.}
\label{fig:corrPl}
\end{figure*}

\end{document}